\documentclass[fleqn,usenatbib]{mnras}

\usepackage{newtxtext,newtxmath}

\usepackage[T1]{fontenc}
\usepackage{array}
\usepackage{booktabs}
\usepackage{orcidlink}
\usepackage{multirow}
\usepackage{makecell}

\DeclareRobustCommand{\VAN}[3]{#2}
\let\VANthebibliography\thebibliography
\def\thebibliography{\DeclareRobustCommand{\VAN}[3]{##3}\VANthebibliography}

\usepackage{graphicx}	
\usepackage{amsmath}	
\usepackage{xcolor}
\usepackage{siunitx}

\newcommand{\lambdaR}{${\lambda_{\mathrm{R}}}$}
\newcommand{\lambdaRe}{${\lambda_{\mathrm{R_e}}}$}
\newcommand{\BTe}{${B/T_\mathrm{e}}$}
\newcommand{\epse}{${\varepsilon_\mathrm{e}}$}
\newcommand{\AgeLW}{${\mathrm{{Age}_{LW}}}$}
\newcommand{\AgeMW}{${\mathrm{{Age}_{MW}}}$}
\newcommand{\Mstar}{$M_{\star}$}
\newcommand{\sSFR}{$\rm{sSFR}_{\rm{R_e}}$}

\title[Morphology--spin connection]{Morphology--spin connection in the SAMI Galaxy Survey}

\author[Baek et al.]
{
Youngmin Baek\orcidlink{0009-0007-5785-0347},$^{1}$\thanks{E-mail: ngc4038@yonsei.ac.kr} 
Sree Oh\orcidlink{0000-0002-4731-9604},$^{1}$\thanks{E-mail: sreemario@gmail.com}
Sukyoung K. Yi\orcidlink{0000-0002-4556-2619},$^{1}$
Scott M. Croom,\orcidlink{0000-0003-2880-9197},$^{2}$
Matthew Colless\orcidlink{0000-0001-9552-8075}, $^{3}$\newauthor
Jesse van de Sande\orcidlink{0000-0003-2552-0021}, $^{4}$
Stefania Barsanti\orcidlink{0000-0002-9332-5386}, $^{2}$
and Sam P. Vaughan\orcidlink{0000-0003-2265-7727} $^{5}$
\\
$^{1}$Department of Astronomy and Yonsei University Observatory, Yonsei University, Seoul 03722, Republic of Korea\\
$^{2}$Sydney Institute for Astronomy, School of Physics, University of Sydney, NSW 2006, Australia\\
$^{3}$Research School of Astronomy and Astrophysics, The Australian National University, Canberra, ACT 2611, Australia\\
$^{4}$School of Physics, University of New South Wales, Sydney, NSW 2052, Australia\\
$^{5}$School of Mathematical and Physical Sciences, Macquarie University, NSW 2109, Australia\\
}

\date{Accepted XXX. Received YYY; in original form ZZZ}

\pubyear{2026}

\begin{document}
\label{firstpage}
\pagerange{\pageref{firstpage}--\pageref{lastpage}}
\maketitle

\begin{abstract}                                                                            
\indent
The spin parameter \lambdaRe\ is a proxy for the specific stellar angular momentum of galaxies and is a useful metric for classifying kinematic morphology. This study aims to quantify the relative importance of galaxy properties in explaining \lambdaRe\, using data from the Sydney-AAO Multi-object Integral-field spectrograph (SAMI) Galaxy Survey. We apply partial correlation analysis and partial least squares regression to assess the relative contributions of different parameters in explaining \lambdaRe. We find that morphology indicators, bulge-to-total ratio within one effective radius (\BTe) and ellipticity (\epse), show the strongest correlations with \lambdaRe\ and play a leading role in the regression analysis. This result statistically confirms the established fast-rotator sequence, in which fast-rotating early-type galaxies form a continuous structural and kinematic sequence with spiral galaxies, with \lambdaRe\ decreasing as bulge prominence increases. The light-weighted age (\AgeLW) and stellar mass (\Mstar) also exhibit significant correlations, but their contributions are secondary to the morphology indicators in multivariate analyses. We also examine whether the observed trends in \lambdaRe\ can be reproduced using galaxy properties alone.
The morphology indicators (\BTe, \epse) reproduce the overall distribution of observed \lambdaRe\ with a scatter of about 0.12, while the inclusion of \AgeLW\ and \Mstar\ provides only modest additional improvement. However, these relations do not reproduce the slow-rotator regime well. Overall, our results show that photometric structural parameters best explain \lambdaRe\ and suggest that statistical inference of galaxy spin from non-IFS observables may become feasible with improved models and a broader set of parameters.
\end{abstract}

\begin{keywords}
galaxies: evolution -- galaxies: kinematics and dynamics -- galaxies: structure.
\end{keywords}


\section{Introduction} 

Since the beginning of extragalactic astronomy, various morphologies of galaxies have been observed. The Hubble sequence provided a framework for discussing galaxy evolution based on their morphology \cite[]{Hubble1926}. As spectroscopic observations began to measure the rotational velocity and dispersion of galaxies, the picture that early-type galaxies (ETGs) are dispersion-dominated systems and late-type galaxies (LTGs) are rotation-supported systems became widely accepted \cite[]{Binney1976, Bertola&Capaccioli1975, Illingworth1977}. This view was broadly consistent with the formation of rotating disc galaxies through tidal torque theory \cite[]{Peebles1969} and the formation of dispersion-supported ETGs via hierarchical mergers \cite[]{Toomre1977, Barnes1992}. 

With the advent of integral-field spectroscopic surveys such as SAURON \cite[]{Bacon2001} and ATLAS3D \cite[]{Cappellari2011a}, spatially resolved studies of galaxy structures and kinematics became possible. These surveys revealed that the majority of ETGs exhibit significant rotation \cite[]{Emsellem2007}. This finding led to the development of a new classification scheme based on the dominance of rotation or random motion, often referred to as 'kinematic morphology', in which galaxies are classified as either fast- or slow-rotators: Fast-rotators are dominated by rotation, slow-rotators are dominated by random motion. The ATLAS3D survey found that 224 out of 260 ETGs are fast-rotators \cite[]{Emsellem2011}, showing that early-type morphology does not necessarily imply a dispersion-dominated kinematic state.

The spin parameter (\lambdaR) is defined as a proxy for the projected stellar angular momentum of galaxies \cite[]{Emsellem2007} and quantifies the ratio of ordered to random motion. The position of a galaxy on the $\lambda_{\mathrm{R}}$--$\varepsilon$ diagram is used to distinguish slow- and fast-rotators, and it separates these populations more clearly than $V/\sigma$ based diagnostics \cite[]{Emsellem2007, Emsellem2011, Cappellari2016, Sande2021a}. Beyond visual morphology, the kinematics of galaxies can now serve as a robust classification metric. 

Many previous studies have shown that the galaxy spin parameter $\lambda_{\mathrm{R}}$  (or angular momentum) 
is correlated with a wide range of galaxy properties. Notably, morphology has been quantitatively identified as an indicator of galaxy spin. According to \cite{Cortese2016}, galaxy specific angular momentum is correlated with stellar mass, and the scatter in this relation is correlated with visual morphology, Sérsic index, and concentration index. Similarly, the relationship between angular momentum, mass, and bulge fraction in spiral galaxies was identified \cite[]{Obreschkow2014}, and later extended by \cite{Sweet2018}, who demonstrated that this relation depends on whether the bulge is classical or pseudo. Most importantly, \citet{Tabor2019} and \citet{Oh2020} performed spectroscopic bulge--disc decomposition based on photometrically derived weights, and found that the bulge and disc components exhibit distinct kinematic properties. These findings are consistent with the established picture that fast-rotating early-type galaxies form a continuous structural and kinematic sequence with spiral galaxies, with galaxy spin decreasing as bulge prominence increases (e.g. \citealt{Cappellari2026} and references therein). Within this framework, the bulge-to-total ratio $(B/T)$ is expected to be a key structural parameter associated with \lambdaR, and clarifying the $B/T$--spin connection may therefore provide insight into the underlying mechanisms of spin evolution.

While morphology plays a significant role, the spin parameter also shows correlations with environment, mass, and age. Analogous to the morphology--density relation \cite[]{Dressler1980}, \cite{Cappellari2011b} suggested a kinematic morphology--density relation which reveals a higher fraction of slow-rotators in denser local environments. However, in studies with larger samples from multi-object integral-field spectroscopic surveys such as Sydney-AAO Multi-object Integral-field spectrograph (SAMI, \citealt{Croom2012}) and Mapping Nearby Galaxies at Apache Point Observatory (MaNGA, \citealt{Bundy2015}), the relationship between spin and local environment is subtle. \cite{Brough2017} and \cite{Greene2017} proposed that stellar mass is the primary driver of the kinematic morphology--density relation because there is no significant correlation between the spin parameter and local environment density at fixed stellar mass. However, it has been suggested that the role of environment in shaping the spin parameter remains even after stellar mass is accounted for \cite[]{Sande2021b}. The most recent studies suggest that the galaxy spin parameter is more correlated with light-weighted age than mass or environment \cite[]{Croom2024}, and that the environment does not provide any further information about the spin parameter when stellar mass, effective radius ($R_{\rm{e}}$), star formation rate (SFR), and ellipticity ($\varepsilon$) are considered \cite[]{Vaughan2024}. Expanding beyond local environment, \cite{Barsanti2025} found that the large-scale environment (specifically the distance to filaments), while secondary to stellar mass and age, shows a stronger correlation with the spin parameter than local environment.

Therefore, in this study we use the SAMI Galaxy Survey as the primary sample to quantify the relative importance of galaxy properties in explaining \lambdaRe, with particular focus on aperture-matched bulge fraction \BTe, which has not been systematically compared with stellar age, stellar mass, and environment in this context. We therefore employ partial correlation analysis and partial least squares regression to assess the relative importance of these parameters across the full sample. To assess the robustness and generalizability of the conclusions drawn from the SAMI sample, we perform a consistency check using the MaNGA \cite[]{Bundy2015} sample in Appendix~\ref{sec:MaNGA_consistency}.
We also aim to assess the feasibility of estimating the spin parameter, which requires spatially resolved kinematics from IFS observations, using only galaxy parameters. This approach can provide a cost-effective way to infer the galaxy kinematic state, enabling spin studies to be extended to much larger samples than are currently possible. Throughout this work, we assume a cosmology with $\Omega_{\mathrm{m}}\!=\!0.3$, $\Omega_{\Lambda}\!=\!0.7$, and $H_0\!=\!70\ \mathrm{km\ s^{-1}\ Mpc^{-1}}$.

\section{Data}
\subsection{SAMI Galaxy Survey}
The SAMI Galaxy Survey is a large spectroscopic survey designed to study the spatially resolved properties of galaxies in the local universe \cite[]{Croom2012}. One of the main scientific goals of the SAMI survey is to understand how angular momentum (also focus of this study) is built up within galaxies. It utilizes the SAMI instrument mounted on the Anglo-Australian Telescope, which employs 13 hexabundles, each containing 61 fibres \cite[]{Bryant2015}. SAMI Galaxy Survey Data Release 3 \cite[]{Croom2021} covers more than 3,000 galaxies spanning a broad range of stellar masses ($8<\log(M_\star/M_\odot)<12$), environments, and morphological types, making it suitable for statistical studies of galaxies. The survey targets were drawn from both the GAMA survey regions \cite[]{Driver2011} and eight galaxy clusters \cite[]{Owers2017}. In this work, we combine the SAMI DR3 products with a set of value-added catalogues from the SAMI collaboration and the literature. Below we describe the adopted spin parameter measurements and the ancillary data used in our analysis.

\subsection{Spin parameter \lambdaRe}
\label{sec:spin}
The spin parameter \lambdaR\ is defined as a proxy for the stellar angular momentum of galaxies \cite[]{Emsellem2007}:
\begin{equation}
\lambda_{\rm{R}} \equiv \frac{\langle R\vert V\vert\rangle}{\langle R \sqrt{V^2+\sigma^2}\rangle}
=
\frac{\Sigma_i F_iR_i\vert V_i\vert}{\Sigma_i F_iR_i\sqrt{V_i^2+\sigma_i^2}}
\label{eq:quadratic}
\end{equation}
where $F_i, V_i, \text{and } \sigma_i$ are the flux, velocity, and velocity dispersion of each spaxel, respectively. In this study, we analyse the spin parameter within $R_{\rm{e}}$, hereafter \lambdaRe, and examine its relation to various galaxy properties.
 
We adopt \lambdaRe\ values from the samiDR3Stelkin catalogue \cite[]{Sande2021a,Sande2021b}, in order to maintain consistency with other SAMI studies that use the same kinematic pipeline and quality definitions. In that work, stellar kinematics were measured using the penalized pixel-fitting (pPXF) method with the MILES stellar library to derive optimal templates for each spaxel, applying the following spaxel quality criteria: signal-to-noise ratio ($\mathrm{S/N}$)$>3$\,\AA$^{-1}$, $\sigma_{\mathrm{obs}}>35\,{\mathrm{km\ s^{-1}}}$, $V_{\mathrm{err}}<30\,{\mathrm{km\ s^{-1}}}$, and $\sigma_{\mathrm{err}}<\sigma_{\mathrm{obs}}\times0.1+25\,{\mathrm{km\ s^{-1}}}$ \cite[]{Sande2017a}. When the kinematic coverage within $R_{\rm{e}}$ is incomplete, an aperture correction is applied. Since kinematic measurements are unreliable for low-mass galaxies due to limited spectral resolution and low S/N, we select galaxies with $M_\star>10^{9.5}M_\odot$. 

Beam smearing corrections are often applied to \lambdaRe\ in IFS surveys in order to mitigate distortions in observed kinematic measurements caused by seeing. The SAMI team catalogue also provides \lambdaRe\ values with an empirical beam-smearing correction based on simulated data, using the Sérsic index and the seeing conditions of individual galaxies \cite[e.g.][]{Harborne2020}. We adopt beam-smearing-corrected \lambdaRe\ in the main results because such corrections are intended to recover the intrinsic kinematics of galaxies. We note that our sample is restricted to galaxies with $\sigma_{\rm PSF}/R_{\rm e} < 0.5$ (Section~\ref{sec:Sample_selection}), which limits the impact of beam smearing on the measured kinematics. Since this study quantifies the correlation between \lambdaRe\ and morphology indicators, including \epse\ and \BTe\ (which correlates with the Sérsic index), applying a morphology dependent correction to the dependent variable could introduce bias in the inferred correlations. Because the beam smearing correction \cite[]{Harborne2020} depends on $\sigma_{\rm PSF}/R_{\rm maj}$ and also on the observed ellipticity and Sérsic index. To demonstrate that this choice does not affect our main qualitative trends and conclusions, we repeat the key analyses using seeing-affected \lambdaRe\ (i.e. beam-smearing correction is not applied) and present the results in Appendix~\ref{sec:Beam_smearing}.

\subsection{Ancillary data}
Structural parameters such as the position angle (PA), effective radius ($R_{\rm{e}}$), and apparent ellipticity ($\varepsilon_{\mathrm{e}}$) were taken from the MGEPhotomUnregDR3 catalogue \cite[]{D'Eugenio2021}, which is based on the multi-Gaussian expansion (MGE; \citealt{Emsellem1994}) fit to the \textit{r}-band photometric images.

Galaxy stellar masses were estimated using the empirical relation between \textit{g–i} colour and \textit{i}-band magnitude \cite[]{Taylor2011}, which assumes a \citet{Chabrier2003} initial mass function (IMF), as implemented in the InputCatGAMADR3 \cite[]{Bryant2015} and InputCatClustersDR3 \citep{Owers2017} catalogues. 

For stellar population properties, we used the values of light-weighted age ($\mathrm{{Age}_{LW}}$) and mass-weighted age ($\mathrm{{Age}_{MW}}$) measured by \citet{Vaughan2022}. The ages were derived by fitting aperture spectra extracted within 1\,$R_{\rm{e}}$ using the penalized pixel-fitting code (pPXF; \citealt{{Cappellari2004, Cappellari2017}})  with MILES simple stellar population models \cite[]{Sanchez2006}, following the procedure described in \citet{Vaughan2022}. 
The SFR values were taken from the EmissionLine1compDR3 catalogue \citep{Scott2018}, where SFRs were measured by summing the attenuation-corrected H$\alpha$ flux from all spaxels within 1\,$R_{\rm{e}}$ and converting it into an SFR using the calibration of \citet{Kennicutt1994}, adjusted for the \citet{Chabrier2003} IMF. 
We calculated the specific star formation rate within one effective radius as $\mathrm{sSFR_{R_e}}\equiv\mathrm{SFR_{R_e}}/M_\star$, where $M_\star$ is the total stellar mass. We also defined the surface star formation rate density as $\Sigma_{\mathrm{SFR}}\equiv\mathrm{SFR_{R_e}}/R_{\rm e,circ}^2$, where $R_{\rm e,circ}$ is the circularized effective radius.

For the local environment metric, we adopted the 5th-nearest-neighbour surface density ($\Sigma_5$) following \citet{Brough2017}. This quantity is estimated from the projected comoving distance to the 5th-nearest galaxy within a redshift window of $\pm1000\ \mathrm{km\ s^{-1}}$, using galaxies from the GAMA survey \cite[]{Driver2011} and the SAMI Cluster Redshift Survey \cite[]{Owers2017}.

Visual morphology classifications were obtained from the VisualMorphologyDR3 catalogue \cite[]{Cortese2016}, where the classification was determined through a voting process among SAMI team members. Cases where consensus was not reached were labelled as unmatched.

\section{Measurement and Sample selection}

\subsection{\BTe\ measurements}
\label{sec:BTe_measurements}
Since bulge and disc are kinematically distinct components of galaxies \cite[]{Oh2020}, we expected that the bulge-to-total ratio within effective radius (\BTe) would be correlated with \lambdaRe. For the present work, we made use of the bulge–disc decomposition catalogues provided by \citet{Barsanti2021} and \citet{Casura2022}. These catalogues are based on photometry from the SDSS \cite[]{York2000}, VST/ATLAS \cite[]{Shanks2013, Shanks2015} and KiDS \cite[]{DeJong2017} surveys, and the decomposition was performed with the source finding and image analysis package PROFOUND \cite[]{Robotham2018} and the bayesian 2D galaxy profile modelling code PROFIT \cite[]{Robotham2017}. 

In \citet{Barsanti2021}, each galaxy was fitted with both a single-component Sérsic profile and a two-component bulge+disc model. To determine whether a galaxy is better described by one or two components, a threshold of $\text{ln(BF)}>60$ was adopted. BF represents the Bayes factor, which is used to compare the probabilities of competing models while penalising the number of parameters and thus avoiding overfitting. For galaxies identified as double-component systems, three different bulge+disc models were tested: (i) de Vaucouleurs ($n{=}4$) + exponential ($n{=}1$), (ii) simple Sérsic (free $n$) + exponential, and (iii) Sérsic bulge with free $n$, axial ratio, and PA + exponential disc. The more complex model was selected if it provided a significantly better fit with $\text{ln(BF)}>10$. 

In \citet{Casura2022}, each galaxy was fitted with three candidate models. (i) single-component Sérsic profile, (ii) Sérsic bulge (free $n$, axial ratio, and PA) + exponential disc, (iii) point-source bulge + exponential disc. Model selection was based on differences in the deviance information criterion (DIC), with the selection thresholds calibrated against visual inspection of a representative subsample. Full details of the decomposition procedure are described in \citet{Barsanti2021} and \citet{Casura2022}.

Using the bulge and disc parameters provided in these catalogues, we reconstructed two-dimensional flux maps for each component. We used the available double-component fits to construct the bulge and disc maps for all galaxies, including those whose preferred model was single-component. In our final sample, 1005 galaxies prefer a double-component model, whereas only 26 prefer a single-component model. Because the latter constitute only a small minority of the sample, this choice does not affect our main results. From these flux maps, we measured the bulge-to-total ratio within an aperture of one effective radius. Therefore, \BTe\ in this study represents the bulge fraction measured at the same aperture as \lambdaRe. This differs slightly from the total $B/T$, which is commonly used as an indicator of galaxy morphology. \BTe\ is expected to provide a more direct structural counterpart to the kinematic measurement because both quantities are measured within $R_{\rm{e}}$. In contrast, global $B/T$ includes light from the outer disc beyond $R_{\rm e}$, which may dilute the connection to the inner kinematic state traced by \lambdaRe.

\subsection{Sample selection}
\label{sec:Sample_selection}
We constructed our sample from the SAMI Galaxy Survey. As our aim is to investigate an empirical relation that can generally explain the spin parameter, the sample selection was performed with minimal restrictions while ensuring sufficient data quality (See Table~\ref{tab:Sample selection}). 

Starting from the SAMI DR3 parent sample, we first applied a stellar mass cut of $\log(M_\star/M_\odot)>9.5$ (Step~1; $N=2192$) to ensure reliable kinematic measurements, as described in Section~\ref{sec:spin}. We then required that a robust \lambdaRe\ measurement is available (Step~2; $N=1673$). To further control the impact of spatial resolution on the measured kinematics, we imposed a cut of $\sigma_{\rm{PSF}}/R_{\rm e}<0.5$ (Step~3; $N=1600$).

After these kinematic-quality and resolution cuts, we required the availability of the structural parameters includes \BTe\ and \epse\ measured at one effective radius. At this stage, we also excluded galaxies where the bulge and disc components might have been swapped during the bulge--disc decomposition. Specifically, we removed 7 galaxies that simultaneously satisfied the criteria of a bulge Sérsic index less than 1.5 and a bulge effective radius larger than that of the disc (Step~4; $N=1053$). We further selected galaxies with available light-weighted age (\AgeLW) measured within $1\,R_{\rm e}$ (Step~5; $N=1032$). Finally, to characterise the local environment we adopted the 5th-nearest-neighbour surface density ($\Sigma_5$) measurement (Step~6; $N=1031$). Note that \BTe, \epse, and \AgeLW\ are measured within $1\,R_{\rm e}$, i.e. on the same spatial scale as \lambdaRe, whereas the stellar mass (\Mstar) and the environment metric ($\Sigma_5$) are not.

\begin{figure}
	\includegraphics[width=\columnwidth]{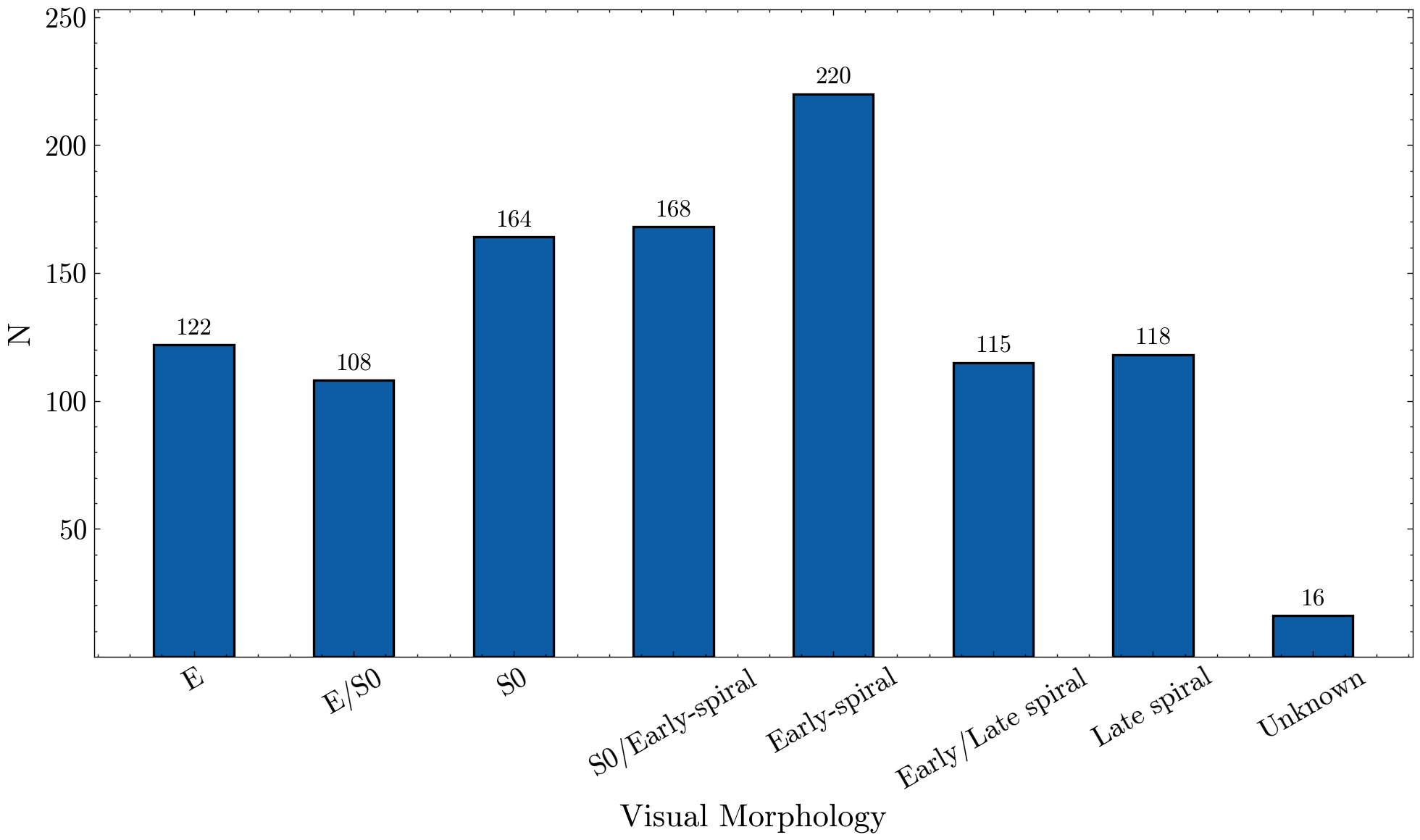}
    \caption{Distribution of visual morphologies in our final sample. Classifications are taken from the VisualMorphologyDR3 catalogue \citep{Cortese2016}, based on SDSS/VST imaging, galaxies without a consensus classification are labelled as Unknown.}
    \label{fig:visual_morph}
\end{figure}

Our final sample spans wide ranges of various galaxy parameters (Fig.~\ref{fig:image1}) and covers a diverse visual morphologies (Fig.~\ref{fig:visual_morph}). Since the SAMI survey includes both GAMA-region and cluster-region targets at $z<0.095$, our final sample (1031 galaxies) contains 656 galaxies from the GAMA regions and 375 galaxies from the cluster regions, providing a broad distribution in environment as traced by $\Sigma_5$.

\begin{table}
\centering
\renewcommand{\arraystretch}{1.2}
\begin{tabular}{c >{\arraybackslash}m{6cm} c}
\toprule
Step & Selection & N \\
\midrule  
(1) & $\log(M_\star/M_\odot)>9.5$ & 2192 \\
(2) & \lambdaRe\ measurement available & 1673 \\
(3) & $\sigma_{\rm{PSF}}/R_{\rm{e}} < 0.5$ & 1600 \\
(4) & \BTe, \epse\ measurement available & 1053 \\
(5) & \AgeLW\ measurement available & 1032 \\
(6) & $\Sigma_5$ measurement available & 1031 \\

\bottomrule
\end{tabular}
\caption{
Step-by-step summary of the sample selection applied in this study.
Starting from the SAMI DR3 parent sample, we apply a stellar mass cut of $\log(M_\star/M_\odot)>9.5$ to ensure reliable kinematic measurements, followed by a series of quality and data-availability cuts on \lambdaRe, spatial resolution ($\sigma_{\rm{PSF}}/R_{\rm{e}}$), and the availability of structural, stellar population, and environmental parameters. The final sample consists of 1031 galaxies.
}
\label{tab:Sample selection}
\end{table}

\section{Method}

To investigate which galaxy properties are most closely linked to the spin parameter \lambdaRe, we employ three complementary statistical approaches. 
First, we use partial correlation analysis \cite[]{Lawrance1976, Croom2024, Barsanti2025} to quantify the association between \lambdaRe\ and galaxy properties while controlling for other variables. 
Second, we apply partial least squares (PLS) regression \cite[]{Geladi1986, Oh2022} to build a multivariate linear model of \lambdaRe\ and to quantify the relative contribution of each property. 
Finally, we use a Random Forest Classifier \cite[]{Breiman2001, Bluck2022, Barsanti2023} to identify which galaxy properties are most informative for separating slow- and fast-rotators.

\subsection{Partial correlation analysis}
In this study, we quantified the correlation between the spin parameter \lambdaRe\ and galaxy properties by calculating the Pearson correlation coefficient and the Spearman rank correlation coefficient. Furthermore, to identify the parameters most strongly associated with the spin parameter \lambdaRe, we performed partial correlation analysis, a statistical method that measures the correlation between two variables while controlling for the effects of other variables. For example, given variables \textit{X}, \textit{Y}, and \textit{Z}, we regress \textit{X} and \textit{Y} on \textit{Z} to obtain residuals, and then calculate the correlation coefficient between these residuals, which yields the pure \textit{X--Y} relationship independent of \textit{Z}. This approach allows us to clarify the direct associations between variables and to identify spurious correlations. The analysis was implemented using \texttt{Pingouin} \cite[]{Vallat2018} which is a statistical package written in Python. 

\subsection{PLS regression}
As an initial step, we examined the linear relationship between each galaxy property and \lambdaRe\ by performing simple linear regression. This allowed us to assess how strongly each parameter individually correlates with the spin parameter before applying a multivariate regression. We implemented the fitting using the \texttt{curve\_fit} function from the \texttt{scipy.optimize} module in Python. 

To extend this analysis, we employed partial least squares (PLS) regression, a multivariate technique particularly suitable when parameters are highly collinear (see Table~\ref{tab:multicollinearity}) or high-dimensional. Similar to principal component analysis (PCA), PLS constructs latent components as linear combinations of the variables; however, unlike PCA, which maximizes the variance of the independent variables alone, PLS extracts components that maximize the covariance between the independent variables and the dependent variable, here \lambdaRe. 

A key advantage of PLS regression for our purposes is that it provides a direct measure of how much variance in \lambdaRe\ is explained by each parameter, allowing us to quantify their relative contributions to the spin parameter. This capability goes beyond assessing correlations, as it highlights which galaxy properties are most influential in explaining variations in \lambdaRe. In addition, we also used PLS regression to explore a predictive model of \lambdaRe, enabling us to evaluate how well galaxy properties can reproduce the spin parameter. The optimal number of PLS components was chosen by minimizing the mean squared error (MSE) between predicted and observed values of \lambdaRe. We implemented this analysis using the \texttt{PLSRegression} class from the \texttt{scikit-learn} package \citep{scikit}, following the implementation described in \citet{Oh2022}.

\subsection{Random Forest Classifier}
\label{sec:RandomForestClassifier}
We employed a Random Forest Classifier to compare and analyse the candidate drivers of galaxy spin reported by previous studies — mass, age, and environment — alongside morphology (\BTe, \epse). Galaxy properties are often correlated and exhibit nonlinear interactions, so simple regression can miss underlying relationships (or causality). A Random Forest addresses these limits by learning nonlinear patterns and interactions, and it remains reliable when features are correlated with each other. It also provides feature importances that indicate which properties contribute most to the classifications. We interpret these values as suggestive signals of possible main drivers of spin. Technically, this analysis predicts kinematic morphology (slow- versus fast-rotator) rather than the continuous spin parameter.

Conceptually, a Random Forest is a collection of many simple decision trees that vote on the final class. Each tree is trained on a bootstrap resample of the data, and at each split it considers only a random subset of features. Each tree learns a sequence of decision criteria that split the data into purer nodes. During training, the algorithm chooses splits that most reduce Gini impurity, which increases node purity (i.e., toward nodes that contain only slow- or only fast-rotators). This randomness reduces similarity among trees and helps prevent overfitting.

Before training, features were transformed to log units and then median centred (following \citet{Barsanti2023} recommendation). Hyperparameters were tuned with \texttt{GridSearchCV} using stratified 5-fold cross-validation and ROC-AUC as the selection metric. ROC-AUC is a measure of discriminative performance of a classifier. We explored the following hyperparameter space: \(n_{\mathrm{estimators}}\in\{50,150,300\}\), \(\mathrm{max\_depth}\in\{\mathrm{None},10,20\}\), \(\mathrm{min\_samples\_leaf}\in\{1,5,15,30\}\), with \(\mathrm{max\_features}=\text{`sqrt'}\), \(\mathrm{class\_weight}\in\{\text{None},\text{`balanced'}\}\), and \(\mathrm{bootstrap}=\text{True}\). The model with the highest ROC-AUC was used for reporting and for assessing feature importance. 

To provide a baseline for interpreting feature importances, we additionally included a random-number feature. This random-number feature carries no physical information about galaxy properties, its measured importance represents the level expected from chance. We therefore interpret only those features with importances clearly exceeding that of the random-number feature as meaningfully informative for the classification.

\section{Result}

\subsection{Correlation of \lambdaRe\ with various parameters}

\begin{figure*}
	\includegraphics[width=\textwidth]{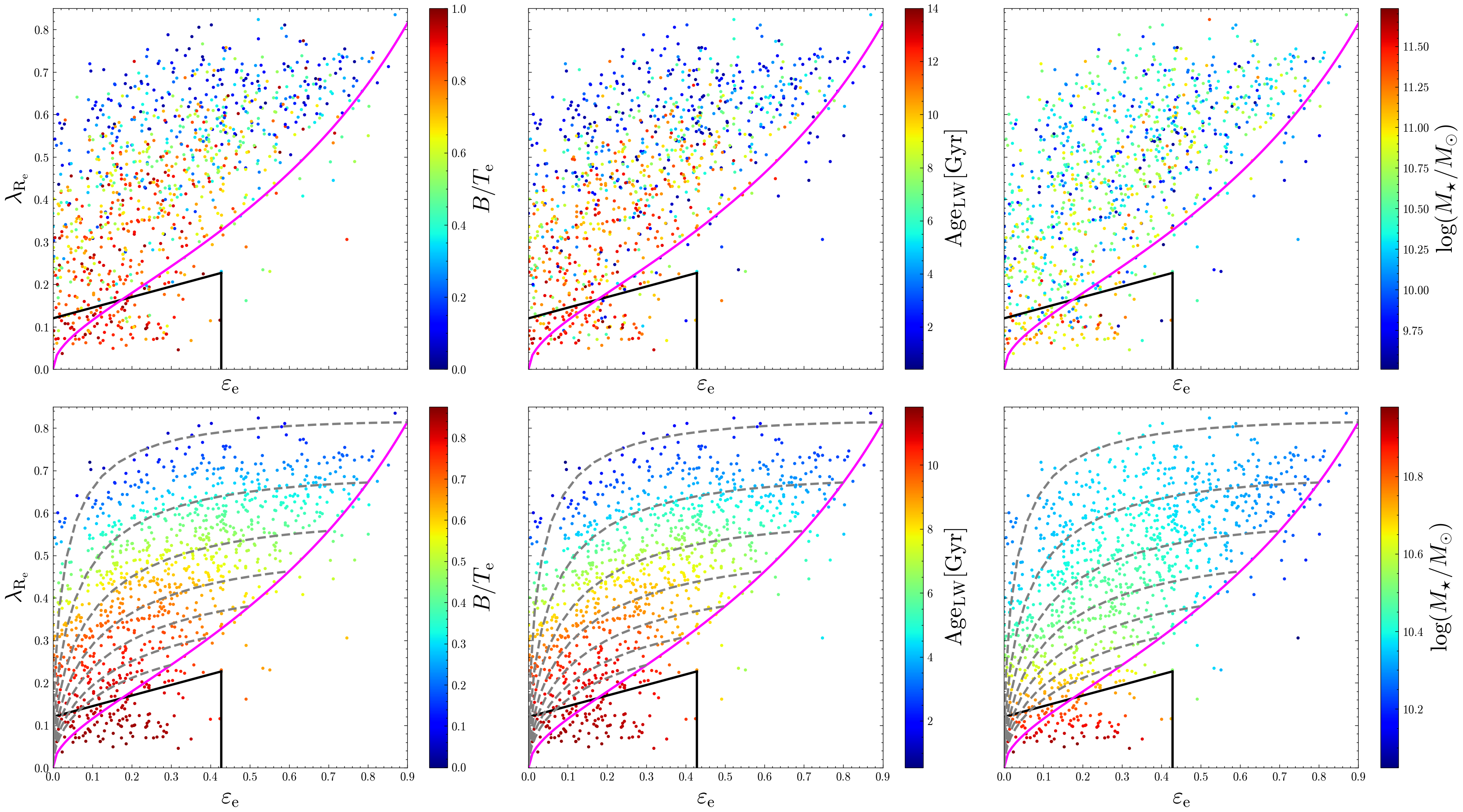}
    \caption[\lambdaRe -- \epse\ diagram]{
    Top row: galaxies in the \lambdaRe -- \epse\ plane, colour-coded by parameters expected to correlate with galaxy spin (\BTe, \AgeLW, $\log(M_\star/M_\odot)$).
    The black trapezoid marks the slow-rotator region, defined following the SAMI-quality data criterion of \citet{Sande2021a}: $\lambda_{\mathrm{R_e}} < 0.12 + \varepsilon_{\mathrm{e}}/4\ \text{and}\  \varepsilon_{\mathrm{e}} < 0.43$.
    Bottom row: LOESS-smoothed \cite[]{Cappellari2013b} versions of the same panels, highlighting the average trends of these parameters across the plane.
    The dashed grey curves show constant intrinsic ellipticity tracks, ranging from 0.3 to 0.9, for axisymmetric systems, with position along each curve corresponding to different inclinations (adopted from \citet{Sande2018}).} 
    \label{fig:lambda-eps-loess}
\end{figure*}

To visually examine how \lambdaRe\ correlates with other galaxy properties, we first plot the \lambdaRe--\epse\ diagram colour-coded by \BTe, \AgeLW, and $M_\star$ (Fig.~\ref{fig:lambda-eps-loess}). We find that \BTe\ varies smoothly across the plane and shows a strong correlation with the spin parameter, comparable to \AgeLW, which has been identified as one of the parameters most strongly correlated with \lambdaRe\ in previous studies \cite[]{Sande2018, Croom2024}. In contrast, $M_\star$ appears to exhibit a dichotomy roughly aligned with the slow-rotator boundary, rather than a strong correlation with \lambdaRe.

\begin{figure*}
	\includegraphics[width=\textwidth]{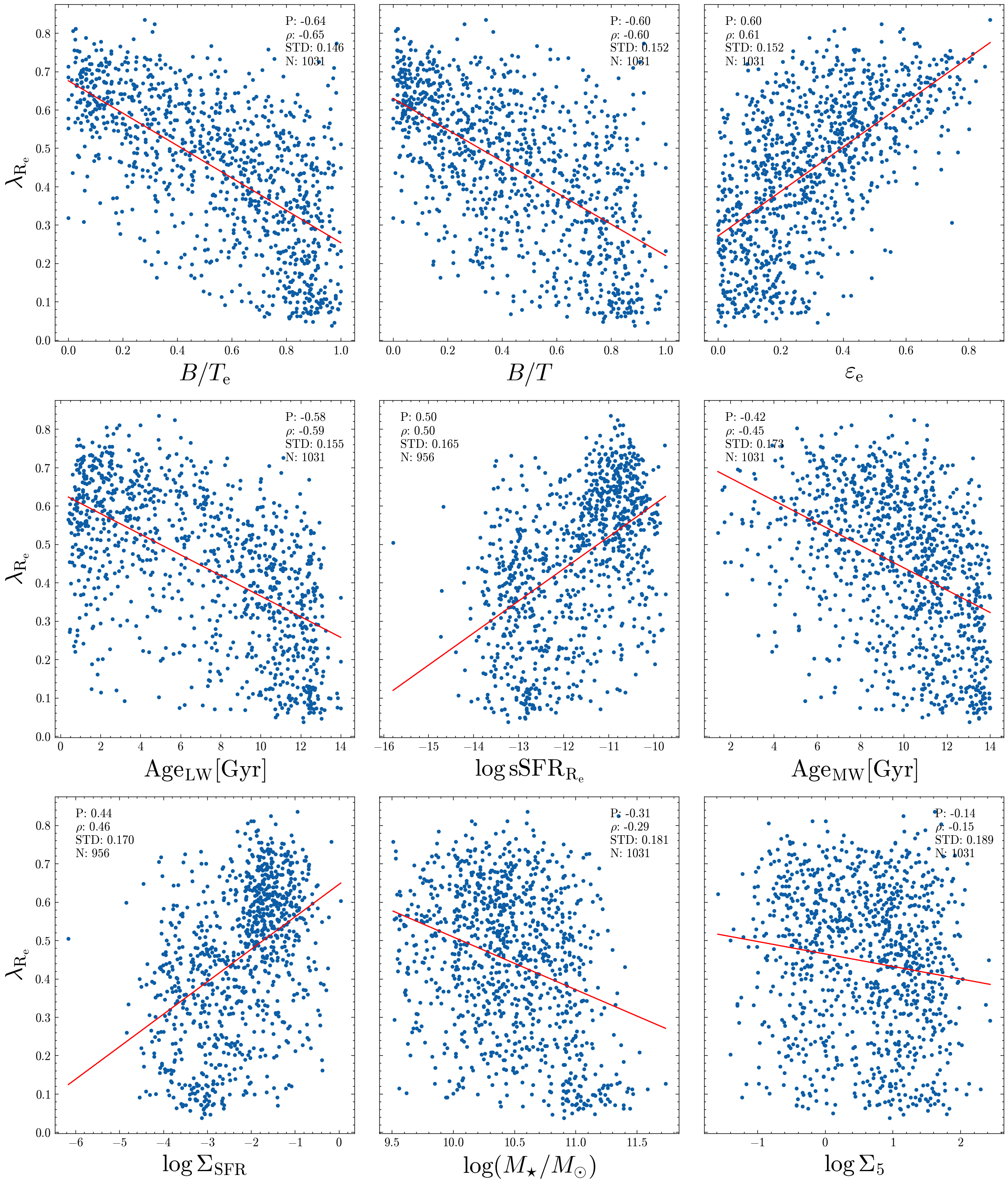}
    \caption[\lambdaRe -- galaxy parameters correlation]
    {
    Plots of spin parameter \lambdaRe\ and various galaxy properties. The x-axis is in order of \BTe, $B/T$, ellipticity (\epse), light-weighted age (\AgeLW), specific star formation rate within effective radius ($\mathrm{sSFR_{R_e}}$), mass-weighted age ($\text{Age}_\text{MW}$) and surface star formation rate density ($\Sigma_{\rm{SFR}}$), stellar mass ($\log(M_\star/M_\odot)$), local environment surface density ($\log\Sigma_5$). Except for $B/T$, stellar mass, and $\Sigma_5$, all galaxy properties are measured within one effective radius. The Pearson correlation coefficient ($P$), Spearman rank correlation coefficient ($\rho$), and standard deviation of the linear fit residual (STD) are noted in the corner of each panel (also summarized in Table~\ref{tab:correlation_coef}).
    } 
    \label{fig:image1}
\end{figure*}

To quantify the correlation with \lambdaRe, in Fig.~\ref{fig:image1}, we present \lambdaRe\ as a function of each parameter, together with a linear fit. In Table~\ref{tab:correlation_coef}, we list the corresponding Pearson and Spearman rank correlation coefficients ($P$ and $\rho$) and the scatter about the fits. Among all parameters considered, the morphology indicators, \BTe\ ($P=-0.64$) and \epse\ ($P=0.60$), show the strongest correlations with \lambdaRe\ and the smallest residual scatter. 
Compared with the total bulge-to-total ratio ($B/T$; $P=-0.60$), \BTe\ shows a slightly stronger correlation, consistent with the expectation that an aperture-matched bulge fraction better traces \lambdaRe. Stellar population parameters also show significant correlations, including light-weighted age (\AgeLW; $P=-0.58$), $\mathrm{sSFR_{R_e}}$ ($P=0.50$), and surface star formation rate density ($\Sigma_{\rm{SFR}}$; $P=0.44$). The mass-weighted age (\AgeMW; $P=-0.42$) shows a weaker correlation than the light-weighted age. The stellar mass (\Mstar; $P=-0.31$) also shows a moderate correlation with \lambdaRe. By contrast, the local environmental surface density ($\Sigma_5$; $P=-0.14$) shows only a weak correlation with \lambdaRe. 

Since our goal is to identify the parameters that are most tightly connected to \lambdaRe\ and best explain its variance, we focus our multivariate analyses on the four parameters (\BTe, \epse, \AgeLW, and \Mstar) which show the strongest correlations. Although other stellar population parameters also exhibit strong correlations, they are tightly correlated with \AgeLW\ (\sSFR; $P=-0.82$, \AgeMW; $P=0.85$, $\Sigma_{\rm{SFR}}$; $P=-0.44$). To avoid redundancy among highly correlated parameters, we retain \AgeLW\ as the representative stellar population indicator, as it shows the strongest correlation with \lambdaRe.

\begin{table}
\centering
\renewcommand{\arraystretch}{1.3}
\begin{tabular}{l c c c c}
\toprule
\textbf{Parameter} & \textbf{$P$} & \textbf{$\rho$} & \textbf{STD} & \textbf{N}\\
\midrule
\BTe              & -0.64 & -0.65 & 0.146 & 1031\\
$B/T$             & -0.60 & -0.60 & 0.152 & 1031\\
\epse             & 0.60  & 0.61  & 0.152 & 1031\\
\AgeLW            & -0.58 & -0.59 & 0.155 & 1031\\
$\log \mathrm{sSFR_{R_e}}$  & 0.50  & 0.50  & 0.165 & 956\\
\AgeMW      & -0.42 & -0.45 & 0.173 & 1031\\
$\Sigma_\mathrm{SFR}$    & 0.44  & 0.46  & 0.170 & 956\\
$M_\star$           & -0.31 & -0.29 & 0.181 & 1031\\
$\log\Sigma_5$   & -0.14 & -0.15 & 0.189 & 1031\\
\bottomrule
\end{tabular}
\caption
{
Correlation coefficients between \lambdaRe\ and various galaxy properties shown in Fig.~\ref{fig:image1}. 
From left to right, the columns list the galaxy property, the Pearson correlation coefficient ($P$), the Spearman rank correlation coefficient ($\rho$), the standard deviation of the linear fit residuals, and the sample size (N).
}
\label{tab:correlation_coef}
\end{table}

Because the above parameters are mutually correlated, the simple Pearson coefficients alone are not sufficient to determine which relations with \lambdaRe\ are truly independent. We therefore use partial correlation analysis to quantify the association of each parameter with \lambdaRe\ after controlling for the remaining variables.
The results of the partial correlation analysis are presented in Table~\ref{tab:partial correlation}. When the other three variables are controlled for, the partial correlations of \AgeLW\ and \Mstar\ with \lambdaRe\ decrease substantially, whereas those of \BTe\ and \epse\ remain relatively strong. For example, the Pearson correlation coefficient between \BTe\ and \lambdaRe\ changes from $P=-0.64$ to $P=-0.42$ when controlling for \epse, \AgeLW, and \Mstar, while the corresponding coefficient for \epse\ changes only from $P=0.60$ to $P=0.50$. In contrast, the partial correlations of \AgeLW\ and \Mstar\ with \lambdaRe\ drop from $P=-0.58$ to $P=-0.27$ and from $P=-0.31$ to $P=0.06$, respectively. These results suggest that \BTe\ and \epse\ are among the parameters most closely associated with \lambdaRe. Notably, \epse\ retains the strongest partial correlation with \lambdaRe\ even when controlling for the others, suggesting that it provides an independent contribution that cannot be replaced by the other parameters. In contrast, \BTe\ is correlated with both \AgeLW\ and \Mstar\ (Table~\ref{tab:multicollinearity}), implying that 
part of the correlations of \AgeLW\ and \Mstar\ with \lambdaRe\ may reflect their correlation with \BTe.

\begin{table}
\centering
\renewcommand{\arraystretch}{1.1}
\scriptsize
\begin{tabular}{c c c c c c}
\toprule
X & Y & Z & Full corr & Partial corr & p-value \\
\midrule
\BTe & \lambdaRe & \epse                         & -0.64 & -0.57 & 2.38E-91 \\
\BTe & \lambdaRe & \AgeLW                        & -0.64 & -0.45 & 9.68E-52  \\
\BTe & \lambdaRe & \Mstar                        & -0.64 & -0.59 & 3.05E-99  \\
\BTe & \lambdaRe & \epse, \AgeLW                 & -0.64 & -0.41 & 1.60E-43  \\
\BTe & \lambdaRe & \epse, \Mstar                  & -0.64 & -0.54 & 1.36E-77  \\
\BTe & \lambdaRe & \AgeLW, \Mstar                 & -0.64 & -0.45 & 2.18E-52  \\
\BTe & \lambdaRe & \epse, \AgeLW, \Mstar          & -0.64 & \textbf{-0.42} & 2.77E-44  \\
\midrule
\epse & \lambdaRe & \BTe                         &  0.60 &  0.52 & 6.12E-73 \\
\epse & \lambdaRe & \AgeLW                       &  0.60 &  0.53 & 2.60E-74 \\
\epse & \lambdaRe & \Mstar                        &  0.60 &  0.58 & 8.66E-94  \\
\epse & \lambdaRe & \BTe, \AgeLW                 &  0.60 &  0.50 & 4.39E-66 \\
\epse & \lambdaRe & \BTe, \Mstar                  &  0.60 &  0.52 & 3.84E-72  \\
\epse & \lambdaRe & \AgeLW, \Mstar                &  0.60 &  0.53 & 2.98E-74 \\
\epse & \lambdaRe & \BTe, \AgeLW, \Mstar          &  0.60 &  \textbf{0.50} & 3.86E-66 \\
\midrule
\AgeLW & \lambdaRe & \BTe                         & -0.58 & -0.31 & 5.24E-25  \\
\AgeLW & \lambdaRe & \epse                        & -0.58 & -0.50 & 4.48E-66  \\
\AgeLW & \lambdaRe & \Mstar                        & -0.58 & -0.52 & 2.87E-71  \\
\AgeLW & \lambdaRe & \BTe, \epse                  & -0.58 & -0.27 & 3.81E-18  \\
\AgeLW & \lambdaRe & \BTe, \Mstar                  & -0.58 & -0.31 & 2.43E-24  \\
\AgeLW & \lambdaRe & \epse, \Mstar                 & -0.58 & -0.45 & 9.85E-52  \\
\AgeLW & \lambdaRe & \BTe, \epse, \Mstar           & -0.58 & -0.27 & 2.44E-18  \\
\midrule
\Mstar & \lambdaRe & \BTe                          & -0.31 & -0.08 & 1.33E-02  \\
\Mstar & \lambdaRe & \epse                         & -0.31 & -0.25 & 5.45E-16  \\
\Mstar & \lambdaRe & \AgeLW                        & -0.31 & 0.00 & 9.43E-01  \\
\Mstar & \lambdaRe & \BTe, \epse                   & -0.31 & -0.05 & 9.53E-02  \\
\Mstar & \lambdaRe & \BTe, \AgeLW                  & -0.31 & 0.06 & 7.39E-02  \\
\Mstar & \lambdaRe & \epse, \AgeLW                 & -0.31 & 0.01 & 8.19E-01  \\
\Mstar & \lambdaRe & \BTe, \epse, \AgeLW           & -0.31 & 0.06 & 5.34E-02  \\
\bottomrule
\end{tabular}
\caption[Partial correlation coefficients table]
{
Results of the partial correlation analysis for \BTe, \epse, \AgeLW, and \Mstar\ (independent variables) with \lambdaRe\ as the dependent variable.
Columns list, from left to right, the independent variable (X), the dependent variable (Y), the controlled variables (Z), the Pearson correlation coefficient between X and Y, the partial correlation coefficient between X and Y controlling for Z, and the corresponding p-value.
}
\label{tab:partial correlation}
\end{table}

\begin{table}
\centering
\renewcommand{\arraystretch}{1.3}
\begin{tabular}{c c c c}
\toprule
\textbf{X} & \textbf{Y} & \textbf{P} & \textbf{p-value}\\
\midrule
\BTe       & \epse       & -0.356 & 3.73E-32  \\
\BTe       & \AgeLW      & 0.609  & 1.64E-105 \\
\BTe       & \Mstar       & 0.404  & 9.69E-42  \\
\epse      & \AgeLW      & -0.345 & 3.29E-30 \\
\epse      & \Mstar       & -0.198 & 1.56E-10 \\
\AgeLW     & \Mstar       & 0.536  & 8.13E-78 \\
\bottomrule
\end{tabular}
\caption[Pairwise correlation between main parameters]
{
Pairwise correlations between the four main parameters analysed in this study (\BTe, \epse, \AgeLW, and \Mstar). Columns list the two correlated variables (X and Y), the Pearson correlation coefficient ($P$), and the corresponding p-value.
}
\label{tab:multicollinearity}
\end{table}

\subsection{Parameter importance from multivariate regression of \lambdaRe}
Based on the correlation analysis in the previous section, we selected \BTe, \epse, \AgeLW, and \Mstar\ as the main predictors of the spin parameter, and derived optimal multivariate linear regression models using PLS regression.

We first calculated the variance proportion in order to quantify the contribution of each parameter to the regression model. 
\begin{equation}
\text{Variance proportion} = \frac{\text{Var}(X_iB_i)}{\text{Var}(Y)}
\label{eq:variance proportion}
\end{equation}
where $X_i$ denotes the independent variables and $B_i$ the corresponding regression coefficients. The variance proportion measures the fraction of the variance in \lambdaRe\ that is explained by each parameter within a given model (Table~\ref{tab:PLS Variance}).

For the model that uses only \BTe\ and \epse, the variance proportions are 0.24 and 0.18, corresponding to 56.8\% and 43.2\% of the explained variance, respectively. When \AgeLW\ is introduced, \epse\ still accounts for 45.9\% of the explained variance and \BTe\ for 39.5\%, whereas \AgeLW\ contributes only 14.6\%. Including \Mstar\ as a fourth parameter leaves the overall pattern unchanged. In that case, \BTe\ contributes 39.0\%, \epse\ 43.8\%, and \AgeLW\ and \Mstar\ contribute 16.7\% and 0.5\%, respectively. These numbers confirm that the explained variance in \lambdaRe\ is dominated by \BTe\ and \epse, while \AgeLW\ plays a secondary role, amounting to about half that of either morphology indicator, whereas \Mstar\ has a negligible effect. This qualitative trend is also reproduced in the MaNGA sample (Appendix~\ref{sec:MaNGA_consistency}).

\begin{table}
\centering
\renewcommand{\arraystretch}{1.2}

\begin{tabular}{ccc}
\toprule
\textbf{Features}         & \BTe & \epse \\ 
\midrule
\textbf{Variance proportion} & 0.24    & 0.18          \\ 
\textbf{Relative contribution}       & {56.8\%}  & {43.2\%}  \\
\bottomrule
\end{tabular}

\bigskip

\begin{tabular}{cccc}
\toprule
\textbf{Features}            & \BTe & \epse & \AgeLW \\ 
\midrule
\textbf{Variance proportion} & 0.13   & 0.16 & 0.05\\ 
\textbf{Relative contribution}          & 39.5\%    & 45.9\% & 14.6\%\\
\bottomrule
\end{tabular}

\bigskip

\begin{tabular}{ccccc}
\toprule
\textbf{Features}            & \BTe & \epse & \AgeLW & \Mstar \\ 
\midrule
\textbf{Variance proportion} & 0.14 & 0.16 & 0.06 & 0.00\\ 
\textbf{Relative contribution}          & 39.0\%    & 43.8\% & 16.7\% & 0.5\%\\
\bottomrule
\end{tabular}
\caption[PLS variance proportion]
{
Each block shows the contribution of each parameter to explaining \lambdaRe\ in the PLS regression model. The variance proportion indicates the ratio of the variance of each feature to the variance of the target variable (\lambdaRe). The relative contribution row represents the fraction of the explained variance attributed to each feature.
}
\label{tab:PLS Variance}
\end{table}

\subsection{Kinematic morphology classification}
For the kinematic morphology classification, using the Random Forest Classifier (Section~\ref{sec:RandomForestClassifier}), we additionally include the 5th-nearest-neighbour surface density ($\Sigma_5$) to enable direct comparison between morphology (the main focus of this study) and other parameters—mass, age, and environment—which have been emphasized as key drivers of spin in previous studies \cite[]{Cappellari2011b, Brough2017, Greene2017, Sande2021b, Croom2024}. We use these parameters in the Random Forest Classifier analysis to quantify their relative importance for determining kinematic morphology, as described in Section~\ref{sec:RandomForestClassifier}.

Using the Random Forest Classifier to predict kinematic morphology (slow-rotator versus fast-rotator; Slow-rotators are defined following the SAMI-quality-data criterion of \citet{Sande2021a}, galaxies satisfying \lambdaRe < 0.12 + \epse/4 for \epse < 0.43). To aid the interpretation of feature importances, we also include a random-number feature, which serves as a null baseline. We therefore interpret physical parameters with importances comparable to the random-number feature with caution. However, we do not interpret this as evidence that environment is physically irrelevant, because $\Sigma_5$ traces only the present-day local surface density and may not capture the merger history, cluster centric location, or large-scale environment that are more directly linked to slow-rotator formation.

As a result, we find that, in the full galaxy sample, \BTe\ has the highest feature importance for distinguishing slow- and fast-rotators (Table~\ref{tab:rf_importance}). In the best model, \BTe\ has the largest importance (0.338), followed by \Mstar\ (0.230), \AgeLW\ (0.167), and \epse\ (0.158), whereas the environmental term ($\log\Sigma_5$) has a much smaller importance (0.060), comparable to that of the random-number feature (0.047). The Top-10 median values across the ROC-AUC ranked models reproduce the same ordering, which indicates robustness to hyperparameter choices.

Since the slow-rotator population is dominated by ETGs, we repeat the same analysis after restricting the sample to ETGs ($N=394$). For the best model, \Mstar\ becomes the most important feature (0.264), while \BTe\ drops to 0.217. \epse\ and \AgeLW\ have comparable but smaller importances (0.160 and 0.153 respectively). $\log\Sigma_5$ remains weak (0.104), close to the corresponding value of the random-number feature. The corresponding Top-10 medians show the same pattern (\Mstar: 0.294; \BTe: 0.227; \epse: 0.161; \AgeLW: 0.154; $\log\Sigma_5$: 0.084; Random feature: 0.075). Therefore, within ETGs the distinction between slow- and fast-rotators becomes more tightly linked to stellar mass than in the full sample, while the influence of morphology indicators is reduced.

\begin{table}
\centering
\small
\renewcommand{\arraystretch}{1.25}
\resizebox{\columnwidth}{!}
{
\begin{tabular}
    {
    l
    S[table-format=1.3]
    S[table-format=1.3]
    S[table-format=1.3]
    S[table-format=1.3]
    }
\toprule
\multirow{2}{*}{\textbf{Features}} &
\multicolumn{2}{c}{\textbf{All galaxies}} &
\multicolumn{2}{c}{\textbf{ETGs only}} \\
\cmidrule(lr){2-3}\cmidrule(lr){4-5}
&
\multicolumn{1}{c}{\textbf{Best model}}
&
\multicolumn{1}{c}{\textbf{Top-10 median}}
&
\multicolumn{1}{c}{\textbf{Best model}}
&
\multicolumn{1}{c}{\textbf{Top-10 median}} \\
\midrule
\BTe             & \textbf{0.338} & 0.342 & 0.217           & 0.227 \\
\epse            & 0.158          & 0.150 & 0.160           & 0.161 \\
\AgeLW           & 0.167          & 0.171 & 0.153           & 0.154 \\
\Mstar           & \textbf{0.230} & 0.226 & \textbf{0.264}  & 0.294 \\
$\log\Sigma_5$   & 0.060          & 0.062 & 0.104           & 0.084 \\
Random feature   & 0.047          & 0.048 & 0.103           & 0.075 \\
\bottomrule
\end{tabular}%
}

\caption[RandomForestClassifier feature importance]
{
Feature importances from Random Forest classifiers for the full galaxy sample ($N=1031$) and the ETG subsample ($N=394$).
The ``Best model'' column corresponds to the model selected by GridSearchCV using ROC-AUC, while the ``Top-10 median'' column gives the median importances across the 10 models with the highest ROC–AUC scores.
}
\label{tab:rf_importance}
\end{table}

\section{Discussion}
In this study, we identified the parameters that most strongly explain the variance in \lambdaRe\ using correlation analysis and partial least squares (PLS) regression. We further employed a Random Forest Classifier to determine which variables most effectively distinguish kinematic morphology. Morphology indicators (\BTe\ and \epse) explain most of the variance in  \lambdaRe, and \BTe\ emerges as the most important predictor of kinematic morphology. Below, we interpret the physical connections that plausibly underlie these correlations. We also assess the feasibility of estimating \lambdaRe\ from photometric morphology indicators.

\subsection{Morphology--spin connection}
The link between morphology and spin is naturally expected. However, IFS surveys have revealed that a large fraction of ETGs host significant rotational components \cite[]{Emsellem2007, Emsellem2011}, indicating that the Early-/Late-Type dichotomy is not sufficient to describe the kinematic state of galaxies.

Theoretically, S\'ersic index $n{\sim}4$ profiles are associated with dispersion-dominated systems formed through violent relaxation \cite[]{LyndenBell1967, Hjorth1991} and dry mergers, while $n{\sim}1$ profiles arise in rotating discs where gas angular momentum is largely conserved. In this context, \BTe\ can be interpreted as a structural proxy that is broadly linked to the dynamical state of galaxy, irrespective of visual morphology type.
Consistent with this view, spectroscopic bulge–disc decompositions from IFS data presented by \citet{Oh2020} (their fig. 13) and \citet{Tabor2019} (their fig. 6) show that bulge components predominantly occupy the slow-rotator region on the $\lambda$--$\varepsilon$ diagram, whereas disc components lie in the fast-rotator region. This suggests that \BTe\ may serve as a powerful proxy for \lambdaRe, consistent with \citet{Krajnovic2013}, who reported a correlation between \lambdaRe\ and $D/T$ based on a smaller sample. Within this framework, a tight \lambdaRe--\BTe\ correlation is anticipated.

Our results are broadly consistent with this picture. \BTe\ shows one of the strongest correlations with \lambdaRe\ and plays a dominant role in the multivariate analysis. At the same time, the \lambdaRe--\BTe\ relation exhibits non-negligible scatter, implying that \BTe\ is a strong first-order proxy for the galaxy kinematic state, but not a complete description of it. It is therefore important to understand what drives the scatter and under what conditions photometric \BTe\ can provide a more accurate estimate of \lambdaRe.

\begin{figure}
    \includegraphics[width=\columnwidth]{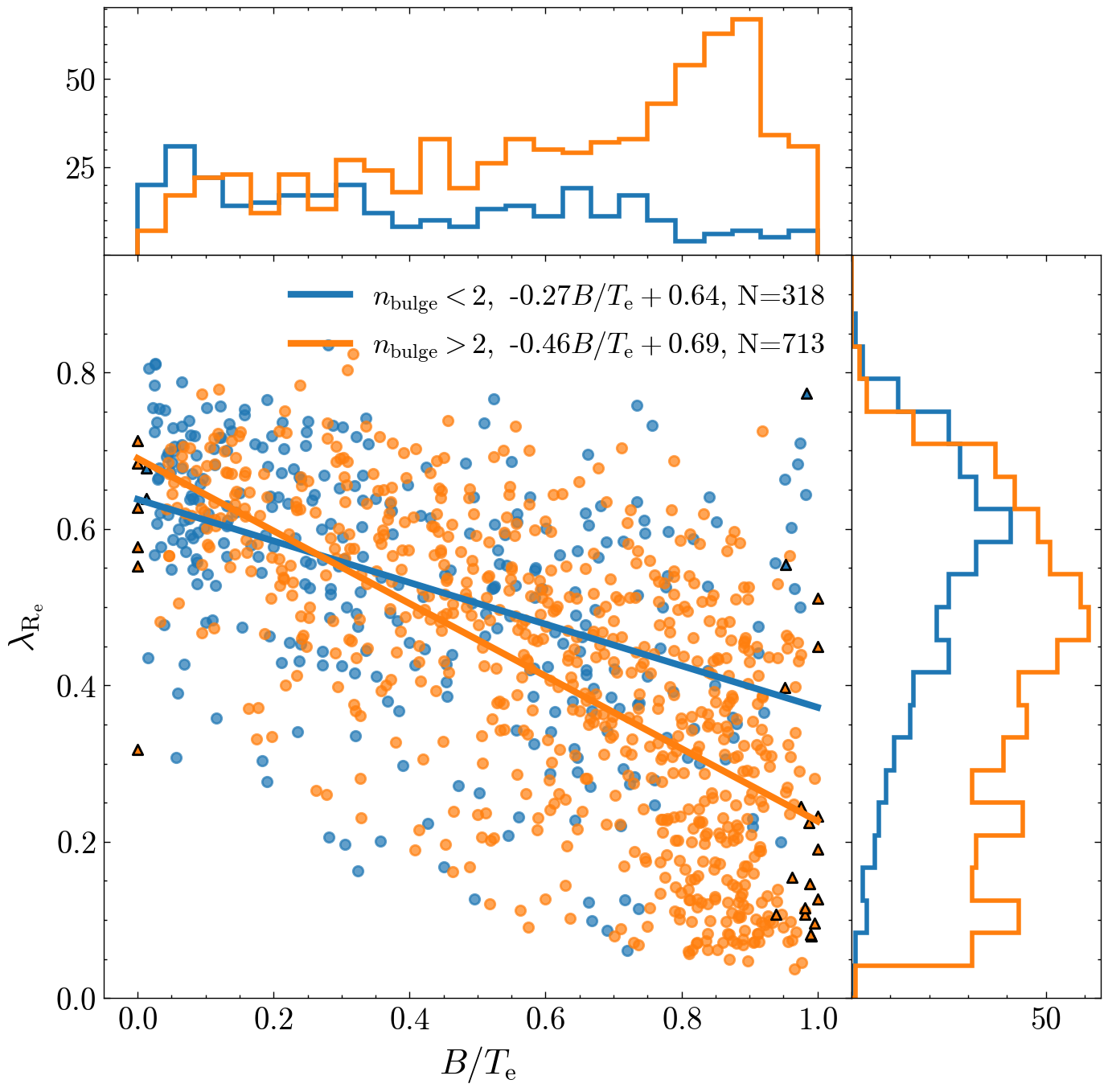}
    \caption[\lambdaRe -- \BTe\ relation by bulge Sérsic index]
    {
    Relation between \lambdaRe\ and \BTe\ for galaxies split by the Sérsic index of the bulge component. Blue points show galaxies with $n_{\rm bulge}<2$ and orange points those with $n_{\rm bulge}>2$. Solid lines indicate the corresponding best-fitting linear relations, with the fitted equations and sample sizes given in the legend. The top and right panels show the histograms of \BTe\ and \lambdaRe\ for each subsample. Galaxies with $n_{\rm bulge}<2$ tend to have higher \lambdaRe\ than those with $n_{\rm bulge}>2$ at fixed \BTe. Triangles with black edges indicate galaxies for which a single-component model is preferred during the bulge--disc decomposition process. 
    }
    \label{fig:classical_pseudo}
\end{figure}

This scatter may be related to bulge type. The relation between \lambdaRe\ and \BTe\ can depend on the Sérsic index of the decomposed bulge component. In Fig.~\ref{fig:classical_pseudo}, we split the sample at $n_{\rm bulge}=2$; the subsample with $n_{\rm bulge}<2$ shows systematically higher \lambdaRe. This suggests that the \BTe--\lambdaRe\ relation depends on bulge type, with pseudo-bulges generally associated with more rotationally supported systems. This agrees with the findings of \citet{Sweet2018}, who reported that the scaling between bulge-to-total ratio and specific stellar angular momentum differs for classical and pseudo-bulges. Accordingly, accurate inference of \lambdaRe\ from \BTe\ may require explicit consideration of bulge type.

We note that bars may also contribute to the scatter in the \BTe--\lambdaRe\ relation. Using the visual bar classifications available for our sample, we find that strongly barred spiral galaxies tend to exhibit lower \lambdaRe\ than non-barred spirals at fixed \BTe\, although we do not attempt a detailed quantitative analysis of this effect here. This trend is consistent with the results of \citet{Joshi2026}, who found that strongly barred galaxies exhibit lower median \lambdaRe\ than non-barred galaxies (their Fig. 12). The presence of a bar may therefore influence the relation between photometric morphology and galaxy spin, in addition to bulge prominence.

Another point to note is that photometric and kinematic $B/T$ are not identical quantities. Using simulated galaxies, \citet{Jang2023} reported a discrepancy between $[D/T]_{\rm phot}$ from mock images and $[D/T]_{\rm kin}$ estimated from orbital circularity. Photometric \BTe\ is luminosity-weighted, whereas kinematic \BTe\ is mass-weighted, so differences are expected. In addition, photometric \BTe\ can be affected by dust extinction and projection. While kinematic \BTe\ should in principle trace the dynamical state more directly, it is worth noting that \lambdaRe\ itself in this study is a luminosity-weighted quantity, which may explain why photometric \BTe\ exhibits a tight correlation with \lambdaRe.

An additional source of the scatter is that the observed ellipticity, \epse, depends on both inclination and intrinsic flattening. Although inclination accounts for much of the variation in \epse\ among fast-rotators, intrinsically flatter systems can reach higher ellipticities and generally occupy the high-\lambdaRe\ regime. Our partial correlation analysis and PLS regression indicate that \epse\ contributes to explaining \lambdaRe\ independently of the other parameters. Some of the intrinsic-flattening dependence is likely already encoded indirectly in structural and stellar-population parameters such as \BTe, \AgeLW, and \Mstar. We therefore interpret the remaining independent contribution of \epse\ as arising primarily from inclination, together with any intrinsic-flattening information not already captured by the other parameters. This complementary role of \epse\ helps explain why combining \BTe\ and \epse\ provides a more complete photometric description of \lambdaRe\ than \BTe\ alone.

\subsection{Spin-down scenarios}
Understanding the formation pathways of slow-rotators is a key objective in studies of galaxy spin evolution. In our analysis, \BTe\ correlates strongly with \lambdaRe\ and is highly informative for kinematic morphology classification. This suggests that the processes linked to bulge growth are also likely to be linked to spin down and, in some cases, to the formation of slow-rotators. We consider two mechanisms for bulge growth: mergers and clumpy disc instability. Mergers are a commonly invoked pathway to classical bulge formation \cite[]{Hopkins2010}. Cosmological simulation studies have shown that dry mergers efficiently reduce galaxy spin \cite[]{Naab2014, Moody2014, Schulze2018}, and \citet{Choi2017} reported that spin-down driven by cumulative minor mergers can exceed that due to major mergers. Clumpy disc instability \cite[]{Noguchi1999, Elmegreen2008, Dekel2009, Ceverino2010} describes a picture in which gas-rich high-redshift discs become gravitationally unstable and form massive clumps that migrate inward to build a dispersion-dominated bulge. Whether the clumps survive long enough for this process remains under debate \citep{Oklopcic2016, Mandelker2016}.

The spin-down scenario favoured by our results is also consistent with an early establishment of the slow-rotator population. The observed slow-rotator fractions remain roughly constant from the local Universe to $z\sim0.75$ \cite[]{Derkenne2024, Mozumdar2025_magnus1}, $19\pm6\%$ in the local MaNGA sample ($z{\sim}0$), $24^{+6.9}_{-4.9}\%$ in the MAGPI survey ($z{\sim}0.3$), and $19.3^{+3.0}_{-2.4}\%$ in the MAGNUS sample ($0.25<z<0.75$). This suggests that the slow-rotator fraction has evolved only weakly over the past several Gyr. If these samples are representative, the weak evolution in the slow-rotator fraction would indicate that a substantial fraction of massive slow-rotators was already in place by $z{\sim}0.75$. This favours spin-down channels that operate efficiently at earlier epochs, such as merger driven bulge growth and possibly clumpy disc instability. While such observational constraints on the formation epoch of massive slow-rotators are promising, current samples remain limited in size and can be sensitive to target selection. Accumulating IFS observations at intermediate to high redshift and developing theoretical predictions from cosmological simulations will therefore be especially valuable. 

\AgeLW\ exhibits a strong correlation with \lambdaRe\ comparable to that of \BTe. This is consistent with the results of \citet{Croom2024}, who analysed stellar mass, stellar population parameters, and environmental metrics using partial correlation analysis, and found \AgeLW\ to be the parameter most strongly correlated with \lambdaRe within that parameter space. However, their analysis did not include structural morphology indicators such as \BTe\ or \epse. When these morphology indicators are included, we find that they provide a stronger explanation of \lambdaRe\ than \AgeLW, while \AgeLW\ remains significant but secondary. We therefore regard the two results as complementary rather than contradictory. \AgeLW\ remains more strongly correlated with \lambdaRe\ than stellar mass and environment in our analysis as well. Part of the difference may also reflect differences in sample construction, since our sample is restricted to galaxies with available \BTe\ measurements.

\citet{Croom2024} proposed secular evolution, progenitor bias, and dry mergers as potential drivers of the \lambdaRe--\AgeLW\ relation. Observationally, the Milky Way shows an age--velocity dispersion relation \cite[]{Nordstrom2004, Sharma2021}, and simulations have suggested that secular dynamical heating can increase velocity dispersion \cite[]{Yi2024}. Bar formation and subsequent secular evolution can redistribute angular momentum and reduce \lambdaRe. \citet{Joshi2026} found that weakly barred systems tend to have younger stellar populations and higher spin than strongly barred systems, including some with boxy/peanut bulges. This suggests that, as secular evolution proceeds, bar growth is associated with increasing \AgeLW, spin-down, and bulge growth.

Progenitor bias is another relevant factor. At high redshift, ionized-gas velocity dispersions are observed to be a factor of $2{\sim}3$ higher than those in the local Universe \cite[]{Wisnioski2015, Ubler2019}, implying that older stellar populations may have been born dynamically hotter. Older galaxies may also be more likely to have experienced dry mergers after quenching, which would further reinforce spin-down and bulge growth \cite[]{Lagos2022}. The physical drivers of the \AgeLW--\lambdaRe\ relation may therefore also contribute to the \BTe\ growth. As shown in Table~\ref{tab:multicollinearity}, \BTe\ and \AgeLW\ are themselves strongly correlated. Nevertheless, \BTe\ provides a better explanation for \lambdaRe\ and for kinematic morphology classification, which may reflect the fact that \BTe\ not only more directly traces the present kinematic state, but may also more closely trace the physical processes responsible for spin-down.

Although stellar mass shows little direct correlation with \lambdaRe\ itself, it contributes substantially to kinematic morphology classification because slow-rotators preferentially reside at the high-mass end (Fig.~\ref{fig:lambda-eps-loess}). This is consistent with a merger-driven picture, since once galaxies become mass quenched, further stellar mass growth increasingly proceeds through mergers, yielding higher ex-situ fractions in the most massive systems \cite[]{Croom2024}. Overall, these results point more naturally to merger-driven bulge growth as the main route to spin down, although secular evolution or clump-driven bulge growth may also contribute in some galaxies.

The secondary role of \AgeLW\ and \Mstar\ can be naturally understood in the context of the (\Mstar, $R_{\rm{e}}$) plane of galaxies. In the framework summarized by \citet{Cappellari2026} (their fig.15 and 16), stellar population properties vary primarily along lines of nearly constant effective velocity dispersion $\sigma_{\rm{e}}\propto(M_\star/R_{\rm{e}})^{1/2}$, which also trace bulge prominence. \AgeLW\ is therefore correlated with \lambdaRe\ partly because both follow the same sequence of increasing bulge prominence, but once \BTe\ and \epse\ are included explicitly, its contribution becomes secondary. Likewise, \Mstar\ alone lacks the size information needed to trace $\sigma_{\rm{e}}$, explaining its weak contribution to the continuous \lambdaRe\ regression. Its larger importance in the ETG-only kinematic morphology classification is consistent with the fact that slow rotators preferentially occupy the high-mass region of the (\Mstar, $R_{\rm{e}}$) plane.

Environmental influence on spin evolution is expected, yet our analysis finds only a weak correlation between $\Sigma_5$ and \lambdaRe, and a minor contribution to the identification of slow-rotators compared to \BTe, \epse, \AgeLW, and $M_\star$. This is consistent with \citet{Vaughan2024}, who found that, after controlling for mass, size, SFR, and ellipticity, $\Sigma_5$ provides no additional predictive power for slow-rotators. Environmental processes undoubtedly matter for galaxy evolution, but the present-day local environment may not fully encode the assembly history. Alternatively, the large-scale environment may hold the key. \citet{Barsanti2025} reported that galaxies closer to filaments tend to have lower spin amplitudes and higher slow-rotator fractions. Spin down driven by mergers within cosmic filaments is physically consistent with our interpretation that bulge growth and spin-down are closely linked. Investigating environmental influences on spin evolution, both local and large-scale, within cosmological simulations will therefore be particularly insightful for advancing our understanding of galaxy spin evolution.

\subsection{Feasibility of empirical inference of \lambdaRe}
\begin{figure*}
	\includegraphics[width=\textwidth]{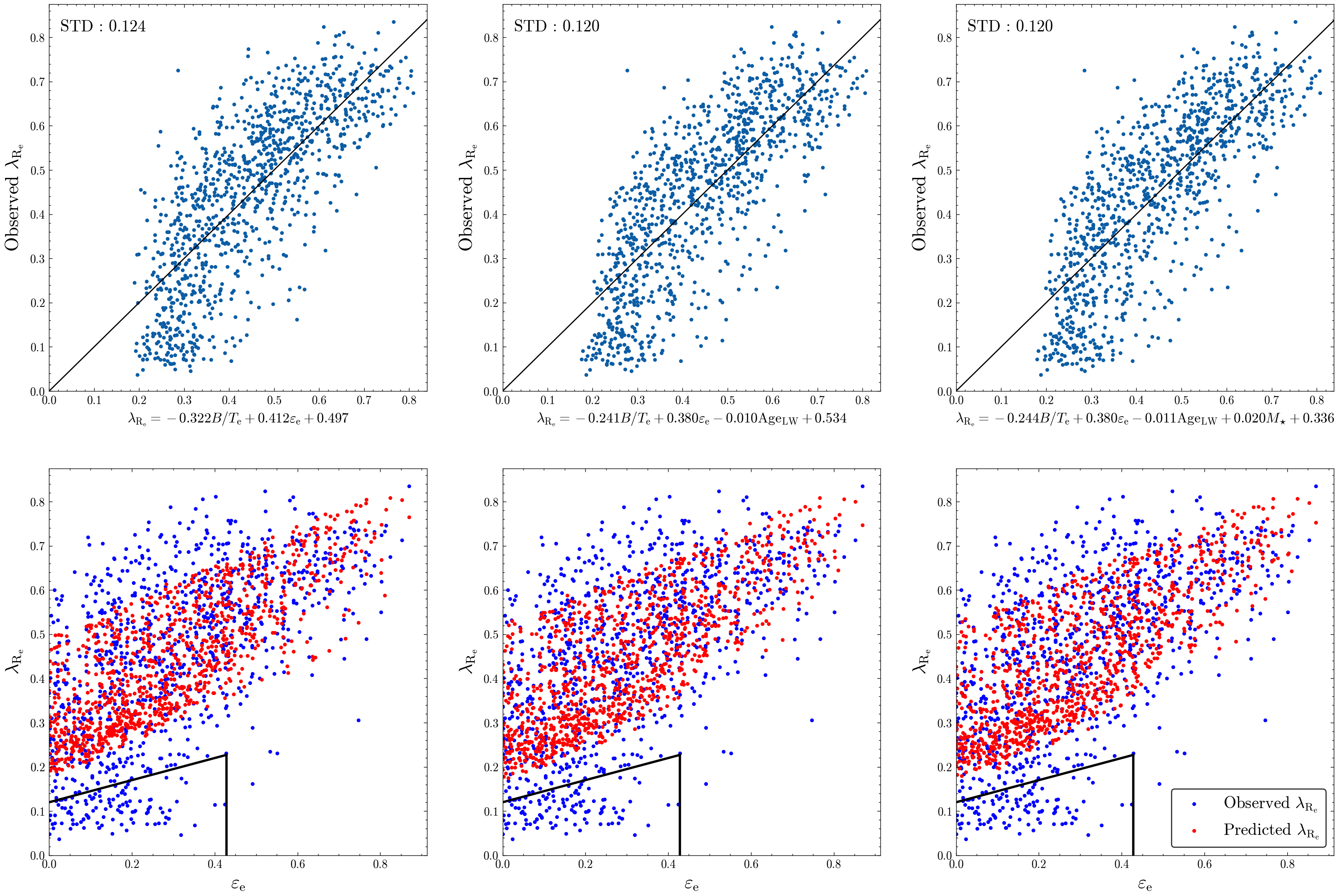}
    \caption
    {    
    Comparison between observed and predicted \lambdaRe. In the top panels, the $x$–axis shows the values of \lambdaRe\ predicted by the PLS regression models and the $y$–axis shows the observed \lambdaRe\ (the one-to-one line is plotted) and the standard deviation of the residuals between predicted and observed values is indicated in the top-left corner of each panel. From left to right, the models use (\BTe, \epse), (\BTe, \epse, \AgeLW), and (\BTe, \epse, \AgeLW, \Mstar) as predictors.
    The bottom panels show the same three models in the \lambdaRe-\epse\ plane, with blue points indicating the observed \lambdaRe\ and red points the PLS predictions, the black trapezoid marks the slow-rotator region.
    }
    \label{fig:PLS_pred}
\end{figure*}

We then evaluated how well the PLS regression models reproduce the observed \lambdaRe. The upper panels of Fig.~\ref{fig:PLS_pred} compare the observed and predicted \lambdaRe\ from the PLS regression model. The corresponding linear relation is indicated at the bottom of each panel. The standard deviation of the residuals between the predicted and observed values is shown in the top left corner. Using two parameters, \BTe\ and \epse, reduces the standard deviation from 0.146 to 0.124 compared to a model with \BTe\ alone (Table~\ref{tab:correlation_coef}). Adding \AgeLW\ further reduces the standard deviation only slightly, to 0.120, and including \Mstar\ yields no improvement. The small change in scatter when \AgeLW\ and \Mstar\ are included is consistent with their relatively small variance proportions.

The PLS regression provides spin estimates with an rms error of $\sim$0.12 using only the photometric parameters \BTe\ and \epse. However, while the regression model captures the overall trend in \lambdaRe, it tends to systematically overestimate \lambdaRe\ for slow-rotators, and consequently does not reproduce the slow-rotator region well (Fig.~\ref{fig:PLS_pred}). 
This limitation is likely related to the established dichotomy between fast and slow rotators. Fast-rotators form a continuous sequence extending from spiral galaxies to discy early-type galaxies, whereas slow rotators represent a distinct population with different intrinsic shapes and evolutionary histories \cite[]{Emsellem2011, Cappellari2016, Sande2021a}. Consequently, a single empirical relation is not expected to fully reproduce the slow-rotator population. We also note that slow rotators become increasingly dominant above a critical mass of $M_\star\sim2\times10^{11}M_\odot$, further supporting the view that they should not be regarded simply as the low-\lambdaRe\ tail of the fast-rotator sequence.

\begin{figure*}
	\includegraphics[width=\textwidth]{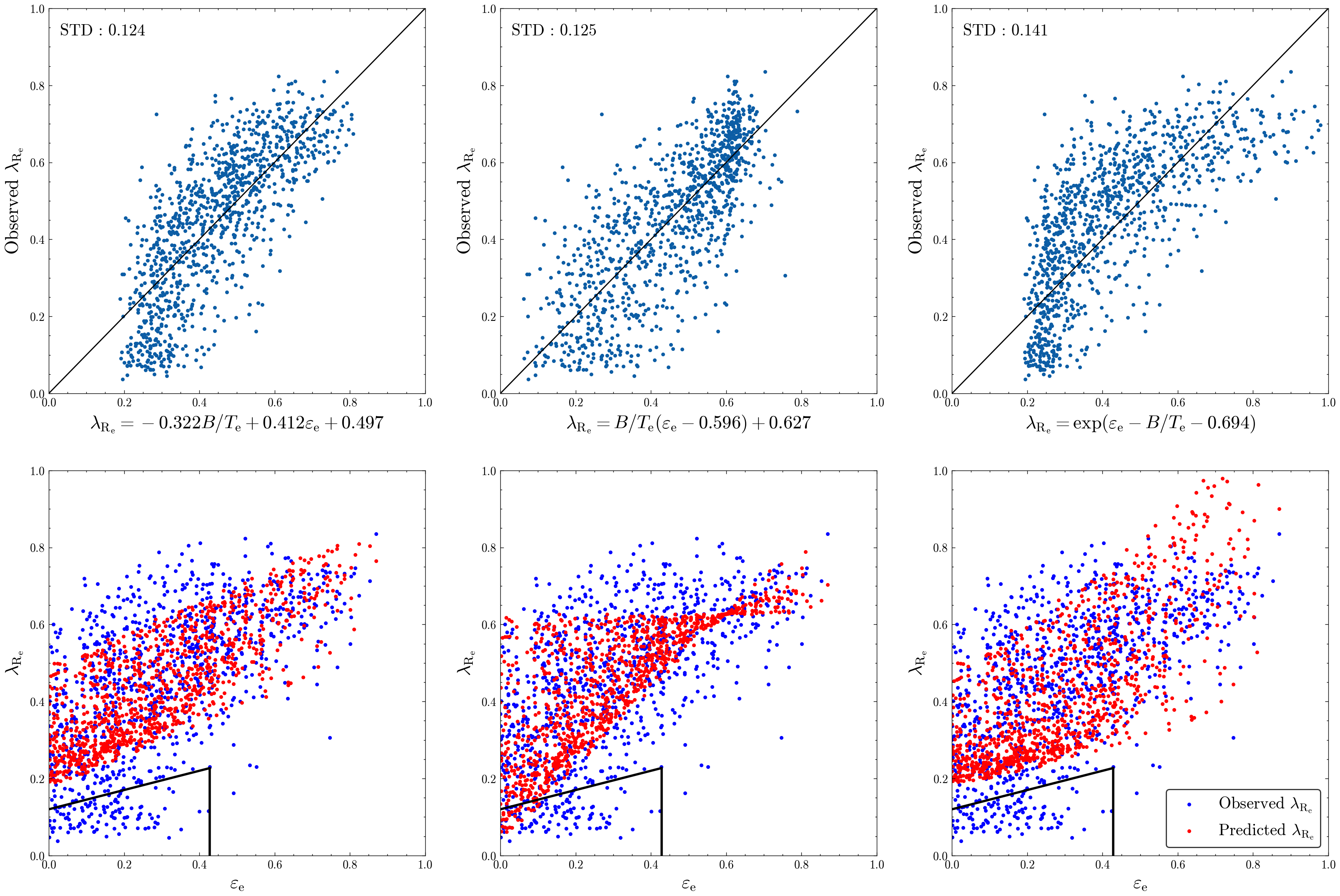}
    \caption
    {Comparison between observed and predicted \lambdaRe\ for the PLS and symbolic regression models. In the top panels, the x-axis shows the values of \lambdaRe\ predicted by each model and the y-axis shows the observed \lambdaRe\ (the one-to-one line is shown). The standard deviation of the residuals between predicted and observed values is indicated in the top-left corner of each panel. From left to right, the three columns show the PLS regression result with \BTe\ and \epse\ (the same as in Fig.~\ref{fig:PLS_pred}), the best symbolic regression model, and the second-best symbolic regression model. The symbolic regression search was performed using \BTe, \epse, \AgeLW, and \Mstar as candidate input variables, and the fitted equations are written below the top panels. The bottom panels show the same three models in the \lambdaRe-\epse\ plane, with blue points indicating the observed \lambdaRe\ and red points presenting the model predictions. The black trapezoid marks the slow-rotator region.
    }  
    \label{fig:symbolic_regression}
\end{figure*}

Nevertheless, the poor reproduction of the slow-rotator regime may also be partly due to the linear form of the PLS model. To test whether a more flexible non-linear relation can better describe the observed distribution, we applied symbolic regression, allowing \BTe, \epse, \AgeLW, and \Mstar\ as candidate input variables. Symbolic regression is a data-driven method that searches over candidate mathematical expressions, rather than assuming a fixed functional form in advance, and is attractive because it can return explicit, potentially interpretable equations \cite[]{Schmidt2009, cranmer2023}. The results are shown in Fig.~\ref{fig:symbolic_regression}. The first column repeats the PLS regression using \BTe\ and \epse\ for direct comparison, while the second and third columns show the best and second-best symbolic regression models, selected by balancing equation complexity and loss. 
The best symbolic regression model improves the description of the slow-rotator regime relative to the PLS fit, but it produces unrealistic distributions at the fast-rotator end, and the resulting equation is difficult to interpret physically.
The second-best model has a simple log-linear form, but the scatter becomes noticeably larger and it again fails to reproduce the slow-rotator regime. Overall, 
allowing more flexible functional forms does not lead to a substantial improvement in predictive performance. At present, we therefore interpret this as indicating that \lambdaRe\ is difficult to describe with a single empirical relation, or that additional parameters not explored here may be crucial for improving predictive performance. 

The relations above are derived from SAMI IFS observations, so their broader applicability should be tested further in other surveys. We do find that the strong $B/T$--\lambdaRe\ relation is also present in the MaNGA sample (Appendix~\ref{sec:MaNGA_consistency}), suggesting that this particular trend is not unique to SAMI. However, the predictive models discussed here have not yet been tested in the same way on external IFS data. In addition, this study is limited to galaxies with $\log{(M_\star/M_\odot)}>9.5$ and $z<0.095$, so the present relations should not be assumed to apply to all galaxies  with available photometry. A next step will be to test these relations in lower-mass galaxies using Hector Galaxy Survey \cite[]{Bryant2024, Oh2025}, which is extending local IFS observations to a larger and more diverse sample. Hector spans $7<\log{(M_\star/M_\odot)}<12$, making it a promising dataset for assessing whether the morphology--spin relation found here remains valid below our current mass limit.

Although the current models do not yet provide accurate estimates of \lambdaRe\ across the full parameter space, the existence of a robust photometric link to \lambdaRe\ suggests that statistical inference of galaxy spin from imaging data may become feasible with improved models and a broader set of parameters, for example two-step formulations that treat fast- and slow-rotators separately, together with physically motivated parameters such as bulge Sérsic index, asymmetry, bar strength, and colour gradients.

\section{Summary and Conclusions}
\label{sec:Conclusions}

In this paper, we investigated which galaxy properties are most closely connected to the stellar angular-momentum proxy \lambdaRe\ using the SAMI Galaxy Survey data, and whether \lambdaRe\ can be statistically linked to galaxy observables. Our results are consistent with the established fast-rotator sequence, in which fast-rotating early-type galaxies form a continuous structural and kinematic sequence with spiral galaxies, from disc-dominated systems to increasingly bulge-dominated discy early-type galaxies. In this context, the dominant role of \BTe\ in explaining \lambdaRe\ provides a statistical confirmation that \lambdaRe\ decreases with increasing bulge prominence along this sequence. Our main conclusions are as follows.

\begin{enumerate}
    \item Morphology indicators, bulge-to-total ratio (\BTe) and ellipticity (\epse), show the strongest correlations with \lambdaRe. The light-weighted age (\AgeLW) and stellar mass (\Mstar) also correlate significantly with \lambdaRe, although less strongly than the morphology indicators.
    
    \item Partial correlation analysis and partial least squares (PLS) regression demonstrate that the morphology indicators are the parameters most strongly correlated with \lambdaRe\ and play the dominant role in explaining its variance. In contrast, the contributions of \AgeLW\ and \Mstar\ are secondary once the morphology indicators are taken into account. We also confirm that this overall trend is broadly consistent with the results based on the MaNGA survey sample (Appendix~\ref{sec:MaNGA_consistency}).

    \item \epse\ provides an independent contribution to explaining \lambdaRe\ that cannot be replaced by the other parameters. We interpret this as evidence that \epse captures both inclination effects and intrinsic flattening that are not fully reflected in the other galaxy properties considered in this work.

    \item Using a Random Forest Classifier to predict kinematic morphology (slow- versus fast-rotator), we find that \BTe\ is the most important feature in the full galaxy sample. When the sample is restricted to ETGs, \Mstar\ becomes the most important feature. 
    In contrast, the environmental surface density ($\Sigma_5$) has an importance comparable to that of a random-number feature, This indicates that $\Sigma_5$ provides little additional predictive power in our models, but should not be taken to imply that environment is physically irrelevant. Rather, $\Sigma_5$ traces only the present-day projected local density and may not capture the cumulative assembly history more directly linked to slow-rotator formation.

    \item Taken together, our results suggest that spin-down is closely linked to bulge growth. Mergers and clumpy disc instabilities are plausible mechanisms for driving this evolution. This scenario is consistent with recent studies at intermediate-redshift ($0.3 < z < 0.7$) that find a slow-rotator fraction comparable to that in the local Universe, implying that a substantial fraction of slow-rotator population was already in place by that epoch.

    \item When we regress \lambdaRe\ on the morphology indicators, the overall distribution of the observed \lambdaRe\ is reproduced with a scatter of about 0.12. However, the \lambdaRe\ values of slow-rotators tend to be overestimated, so the slow-rotator regime is not well reproduced. Despite this limitation, our results suggest that \lambdaRe\ may be statistically linked to non-IFS observables, making statistical studies of galaxy spin feasible in much larger survey samples such as LSST \cite[]{Ivezic2019}.

\end{enumerate}

\section*{Acknowledgements}
This work is based on data from the SAMI Galaxy Survey.  
The SAMI Galaxy Survey is based on observations made at the Anglo-Australian Telescope. 
The survey was supported by the ARC Centres of Excellence CAASTRO (CE110001020) and ASTRO 3D (CE170100013), together with the participating institutions.
The Sydney-AAO Multi-object Integral Field Spectrograph (SAMI) was developed jointly by the University of Sydney and the Australian Astronomical Observatory, and funded by 
ARC grants FF0776384 and LE130100198. 
The SAMI Galaxy Survey website is \url{http://sami-survey.org/}. 
S.K.Y. acknowledges support from the Korean National Research Foundation (RS-2025-00514475; RS-2022-NR070872).
SO acknowledges support from the Korean National Research Foundation (NRF) (RS-2023-00214057).

\section*{Data Availability}
The SAMI Galaxy Survey Data Release 3 products used in this work are publicly available through Australian Astronomical Optics Data Central (\url{https://datacentral.org.au/}). 
The derived data products generated in this work are available from the corresponding author upon reasonable request.


\bibliographystyle{mnras}
\bibliography{references}



\appendix

\section{Consistency check with MaNGA}
\label{sec:MaNGA_consistency}

To examine whether the results obtained from the SAMI sample are generalizable, we performed the same analysis using the Mapping Nearby Galaxies at APO (MaNGA) survey \cite[]{Bundy2015}. The galaxy parameters were drawn from the publicly available Value-Added Catalogs of SDSS-IV MaNGA. Specifically, the MaNGA PyMorph DR17 photometric catalogue \cite[]{Dominguez2022} provided bulge-to-total ratio ($B/T$), bulge axial ratio, and disc axial ratio. The MaNGA Dynamics and Population (DynPop) catalogue \cite[]{Zhu2023_dynpop} provided \lambdaRe, effective radius, velocity dispersion within the effective radius, and ellipticity, and the MaNGA Firefly Stellar Populations catalogue \cite[]{Neumann2022} provided stellar mass, light-weighted age, and mass-weighted age. For consistency with the SAMI sample, we selected galaxies with stellar mass greater than $10^{9.5}\,M_\odot$.  

We first examined the correlation coefficients and found that $B/T$ and \epse\ show the strongest correlations with galaxy spin. A difference arises in the definition of the bulge-to-total ratio. The SAMI analysis employed the bulge-to-total ratio within the effective radius (\BTe), whereas the MaNGA PyMorph catalogue provides a global $B/T$ measured from the overall light distribution. Despite this difference in aperture, both datasets consistently identify $B/T$ as a key parameter correlated with galaxy spin. If a \BTe\ measurement were available for the MaNGA sample, the consistency between the two surveys would likely be even stronger.  

Partial correlation analysis also points to $B/T$ and \epse\ as the primary parameters linked to galaxy spin. Importantly, their correlations remain robust even after controlling for other variables, whereas \AgeLW\ and \Mstar\ lose most of their correlations once control variables are introduced. In addition, PLS regression analysis confirms that $B/T$ and \epse\ contribute the most to explaining the variance in \lambdaRe.  

\begin{table}
\centering
\renewcommand{\arraystretch}{1.4}
\begin{tabular}{l c c c c}
\toprule
\textbf{Parameter} & \textbf{$P$} & \textbf{$\rho$} & \textbf{STD} & \textbf{N}\\
\midrule
$B/T$              & -0.56 & -0.55 & 0.207 & 7927\\
\epse             & 0.57  & 0.59  & 0.204 & 7927\\
\AgeLW            & -0.42 & -0.41 & 0.226 & 7927\\
\AgeMW            & -0.31  & -0.34  & 0.237 & 7927\\
\Mstar            & -0.30 & -0.24 & 0.238 & 7927\\
Bulge axial ratio        & -0.43  & -0.43  & 0.225 & 7927\\
Disc axial ratio          & -0.35 & -0.38 & 0.233 & 7927\\
$\sigma_{\rm{e}}$   & -0.42 & -0.27 & 0.226 & 7927\\
\bottomrule
\end{tabular}
\caption
{
Correlation coefficients between \lambdaRe\ and various galaxy properties in the MaNGA sample. Columns are defined in the same way as in Table~\ref{tab:correlation_coef}.
}
\label{tab:correlation_coef_MaNGA}
\end{table}

\begin{table}
\centering
\renewcommand{\arraystretch}{1.3}
\scriptsize
\begin{tabular}{c c c c c c}
\toprule
X & Y & Z & Full corr & Partial corr & p-value \\
\midrule
$B/T$ & \lambdaRe & \epse, \AgeLW, \Mstar               & -0.56 & -0.40 & 6.97E-295 \\

\epse & \lambdaRe & $B/T$, \AgeLW, \Mstar               &  0.57 &  0.49 & 0.0 \\

\AgeLW & \lambdaRe & $B/T$, \epse, \Mstar               & -0.42 & -0.14 & 2.92E-36  \\

\Mstar & \lambdaRe & $B/T$, \epse, \AgeLW               & -0.30 & -0.02 & 1.58E-01  \\

\bottomrule
\end{tabular}
\caption
{
Partial correlation analysis for $B/T$, \epse, \AgeLW, and \Mstar\ in the MaNGA sample, with \lambdaRe\ as the dependent variable. The columns are defined as in Table~\ref{tab:partial correlation}.
}
\label{tab:MaNGA partial correlation}
\end{table}

\begin{table}
\centering
\renewcommand{\arraystretch}{1.2}

\begin{tabular}{ccc}
\toprule
\textbf{Features}         & $B/T$ & \epse \\ 
\midrule
\textbf{Variance proportion} & 0.18    & 0.20          \\ 
\textbf{Relative contribution}       & {47.6\%}  & {52.4\%}  \\
\bottomrule
\end{tabular}

\bigskip

\begin{tabular}{cccc}
\toprule
\textbf{Features}            & $B/T$ & \epse & \AgeLW \\ 
\midrule
\textbf{Variance proportion} & 0.14    & 0.19 & 0.01\\ 
\textbf{Relative contribution}          & 40.5\%    & 55.3\% & 4.2\%\\
\bottomrule
\end{tabular}

\bigskip

\begin{tabular}{ccccc}
\toprule
\textbf{Features}            & $B/T$ & \epse & \AgeLW & \Mstar \\ 
\midrule
\textbf{Variance proportion} & 0.13    & 0.18 & 0.01 & 0.00\\ 
\textbf{Relative contribution}          & 40.4\%  & 55.3\% & 4.2\% & 0.1\%\\
\bottomrule
\end{tabular}

\caption
{
Same as Table~\ref{tab:PLS Variance}, but for the MaNGA sample, where global $B/T$ is used instead of \BTe.
}
\label{tab:MaNGA PLS Variance}
\end{table}

In summary, although some discrepancies exist due to methodological differences between SAMI and MaNGA (particularly in the definition of $B/T$), the overall analysis yields broadly consistent results. The parameters most closely associated with galaxy spin are the bulge-to-total ratio and ellipticity, indicating that the main conclusions of this study are robust to the choice of IFS survey.

\section{Seeing-affected \lambdaRe}
\label{sec:Beam_smearing}

We repeat the analyses using \lambdaRe\ values without beam-smearing correction. The main qualitative conclusions remain unchanged. Any quantitative differences do not affect the interpretation presented in the main text. 

\begin{table}
\centering
\renewcommand{\arraystretch}{1.4}
\begin{tabular}{l c c c c}
\toprule
\textbf{Parameter} & \textbf{$P$} & \textbf{$\rho$} & \textbf{STD} & \textbf{N}\\
\midrule
\BTe              & -0.68 & -0.67 & 0.133 & 1031\\
$B/T$             & -0.61 & -0.62 & 0.143 & 1031\\
\epse             & 0.63  & 0.62  & 0.140 & 1031\\
\AgeLW            & -0.62 & -0.62 & 0.142 & 1031\\
$\log \mathrm{sSFR_{R_e}}$  & 0.56  & 0.56  & 0.150 & 956\\
\AgeMW      & -0.44 & -0.48 & 0.162 & 1031\\
$\Sigma_\mathrm{SFR}$    & 0.47  & 0.48  & 0.160 & 956\\
\Mstar          & -0.24 & -0.25 & 0.175 & 1031\\
$\log\Sigma_5$   & -0.20 & -0.21 & 0.177 & 1031\\
\bottomrule
\end{tabular}
\caption[Correlation coefficients table (Seeing-affected \lambdaRe)]
{
Correlation coefficients between \lambdaRe\ (seeing-affected) and various galaxy properties. Columns are defined in the same way as in Table~\ref{tab:correlation_coef}.
}
\label{tab:correlation_coef_beam}
\end{table}

\begin{table}
\centering
\renewcommand{\arraystretch}{1.3}
\begin{tabular}{c c c c c c}
\toprule
X & Y & Z & Full corr & Partial corr & p-value \\
\midrule
\BTe & \lambdaRe & \epse, \AgeLW, \Mstar               & -0.68 & -0.50 & 4.16E-65 \\

\epse & \lambdaRe & \BTe, \AgeLW, \Mstar               & 0.63 &  0.57 & 1.67E-88 \\

\AgeLW & \lambdaRe & \BTe, \epse, \Mstar               & -0.62 & -0.38 & 1.53E-37  \\

\Mstar & \lambdaRe & \BTe, \epse, \AgeLW               & -0.24 & 0.26 & 1.75E-17  \\

\bottomrule
\end{tabular}
\caption[Partial correlation coefficients table (Seeing-affected \lambdaRe)]
{
Partial correlation analysis for \BTe, \epse, \AgeLW, and \Mstar\, with \lambdaRe\ (seeing-affected) as the dependent variable. Columns are defined as in Table~\ref{tab:partial correlation}.
}
\label{tab:partial correlation beam}
\end{table}

\begin{table}
\centering
\renewcommand{\arraystretch}{1.2}

\begin{tabular}{ccc}
\toprule
\textbf{Features}         & \BTe & \epse \\ 
\midrule
\textbf{Variance proportion} & 0.27    & 0.20          \\ 
\textbf{Relative contribution}       & {57.2\%}  & {42.8\%}  \\
\bottomrule
\end{tabular}

\bigskip

\begin{tabular}{cccc}
\toprule
\textbf{Features}            & \BTe & \epse & \AgeLW \\ 
\midrule
\textbf{Variance proportion} & 0.15    & 0.17 & 0.06\\ 
\textbf{Relative contribution}          & 39.6\%    & 45.4\% & 15.0\%\\
\bottomrule
\end{tabular}

\bigskip

\begin{tabular}{ccccc}
\toprule
\textbf{Features}            & \BTe & \epse & \AgeLW & \Mstar \\ 
\midrule
\textbf{Variance proportion} & 0.17    & 0.17 & 0.10 & 0.03\\ 
\textbf{Relative contribution}          & 35.4\%  & 36.1\% & 21.8\% & 6.7\%\\
\bottomrule
\end{tabular}

\caption[PLS variance proportion (Seeing-affected \lambdaRe)]
{
Same as Table~\ref{tab:PLS Variance}, with seeing-affected \lambdaRe.
}
\label{tab:PLS Variance beam}
\end{table}

\bsp
\label{lastpage}
\end{document}